\documentclass[aps,prb,twocolumn,superscriptaddress,showpacs]{revtex4}
\usepackage{graphicx}
\usepackage{dcolumn}
\usepackage{bm}
\usepackage{color}
\usepackage{multirow}
\usepackage{hyperref}
\usepackage{amssymb}
\usepackage{color}
\usepackage{graphics}\usepackage{epsfig}
\usepackage{footnote}
\usepackage{amsmath}
\usepackage{multirow}
\usepackage{graphicx}
\usepackage{tabularx}

\begin{document}
\title{On the importance of measuring accurately LDOS maps using scanning tunneling spectroscopy in materials presenting atom-dependent charge order: the case of the correlated Pb/Si(111) single atomic layer}
\author{C.~Tresca}
\affiliation{CNR-SPIN c/o Dipartimento di scienze fisiche e chimiche, Universit\`a degli studi dell'Aquila, Via Vetoio 10, I-67100 L'Aquila, Italy}
\author{T.~Bilgeri}
\author{G.~Menard}
\author{V.~Cherkez}
\author{R.~Federicci}
\author{D.~Longo}
\author{M.~Herv\'e}
\author{F.~Debontridder}
\author{P.~David}
\affiliation{Sorbonne Universit\'e, CNRS, Institut des Nanosciences de Paris, UMR7588, F-75252 Paris, France}
\author{D.~Roditchev}
\affiliation{Sorbonne Universit\'e, CNRS, Institut des Nanosciences de Paris, UMR7588, F-75252 Paris, France}
\affiliation{Laboratoire de physique et d'\'etude des mat\'eriaux, LPEM-UMR8213/CNRS-ESPCI ParisTech-UPMC,
10 rue Vauquelin, F-75005 Paris, France}
\author{G.~Profeta}
\affiliation{CNR-SPIN c/o Dipartimento di scienze fisiche e chimiche, Universit\`a degli studi dell'Aquila, Via Vetoio 10, I-67100 L'Aquila, Italy}
\affiliation{Dipartimento di scienze fisiche e chimiche, Universit\`a degli studi dell'Aquila, Via Vetoio 10, I-67100 L'Aquila, Italy}
\author{T.~Cren}
\affiliation{Sorbonne Universit\'e, CNRS, Institut des Nanosciences de Paris, UMR7588, F-75252 Paris, France}
\author{M.~Calandra}
\email{m.calandrabuonaura@unitn.it}
\affiliation{Sorbonne Universit\'e, CNRS, Institut des Nanosciences de Paris, UMR7588, F-75252 Paris, France}
\affiliation{Dipartimento di Fisica, Universit\`a di Trento, via Sommarive 14, I-38123 Povo, Italy}
\author{C.~Brun}
\email{christophe.brun@sorbonne-universite.fr}
\affiliation{Sorbonne Universit\'e, CNRS, Institut des Nanosciences de Paris, UMR7588, F-75252 Paris, France}

\pacs {75.70.Tj 
 73.20.At, 
 68.37.Ef, 
 71.45.Lr  
}

\begin{abstract}
We address here general issues and show how to properly extract the local charge order in two-dimensional systems 
from scanning tunneling microscopy/spectroscopy (STM/STS) measurements. When the charge order presents spatial variations at the atomic scale inside the unit cell and is energy dependent, particular care should be taken. We show that the widely used lock-in technique performed while acquiring an STM topography in closed feedback loop, cannot be used to extract this local charge order from STS dI/dV differential conductance maps. In such situations, the use of the lock-in technique leads to systematically incorrect dI/dV measurements giving a false local charge order. We show that a correct method is either to perform a constant height measurement or to perform a full grid of dI/dV(V) spectroscopies, using a setpoint for the bias voltage outside the bandwidth of the correlated material where the local density-of-states (LDOS) is expected to be spatially homogeneous. We take as a paradigmatic example of 
two-dimensional material the 1/3 single-layer Pb/Si(111) phase. As large areas of this phase cannot be grown, 
	charge ordering in this system is not accessible to angular resoved photoemission or grazing x-ray diffraction. Two previous investigations by STM/STS supplemented by {\it ab initio} Density Functional Theory (DFT) calculations concluded that this material undergoes a phase transition to a low-temperature $3\times 3$ reconstruction where one Pb atom moves up, the two remaining Pb atoms shifting down. A third STM/STS study by Adler {\it et al.} [PRL 123, 086401 (2019)] came to the opposite conclusion, i.e. that two Pb atoms move up, while one Pb atom shifts down. We show that this latter erroneous conclusion comes from a misuse of the lock-in technique. In contrast, using a full grid of dI/dV(V) spectroscopy measurements, our results show that the energy-dependent LDOS maps agree very well with state-of-the-art DFT calculations confirming the one-up two-down charge ordering. We 
show that this structural and charge re-ordering inside the $3\times 3$ unit cell is equally driven by electron-electron interactions and the coupling to the substrate. 
\end{abstract}

\maketitle

\section{Introduction}

The understanding of the structural and electronic properties of correlated materials represents an important challenge in condensed matter physics. The interplay between Coulomb interactions and electrons delocalization leads to the appearence of new ground states often presenting charge, spin and/or lattice re-ordering \cite{Fazekas1980,Carpinelli1996,Carpinelli1997,Ottaviano2000,PhysRevLett.94.046101,Ghiringhelli2012,Comin2014,Janod2015,PhysRevLett.121.026401,Tresca2019.NbS2}. Probing experimentally these various orders with macroscopic and local probes is essential to provide solid grounds for advanced theoretical modelling. In addition, in the 
goal of reaching a quantitative understanding of the theoretical properties of correlated electronic systems, it is highly desirable to find materials exhibiting the greatest chemical simplicity. 
The ideal situation is that of a small number of atoms in the unit cell possibly leading to a single narrow band  crossing the Fermi level, isolated from other bulk bands and presenting strong on-site Coulomb repulsion. Such a system would ideally implement the single-band Hubbard model at the heart of the description of strongly correlated materials \cite{Hubbard1963}. 

\begin{figure*}
\centering 
\includegraphics[width=\linewidth]{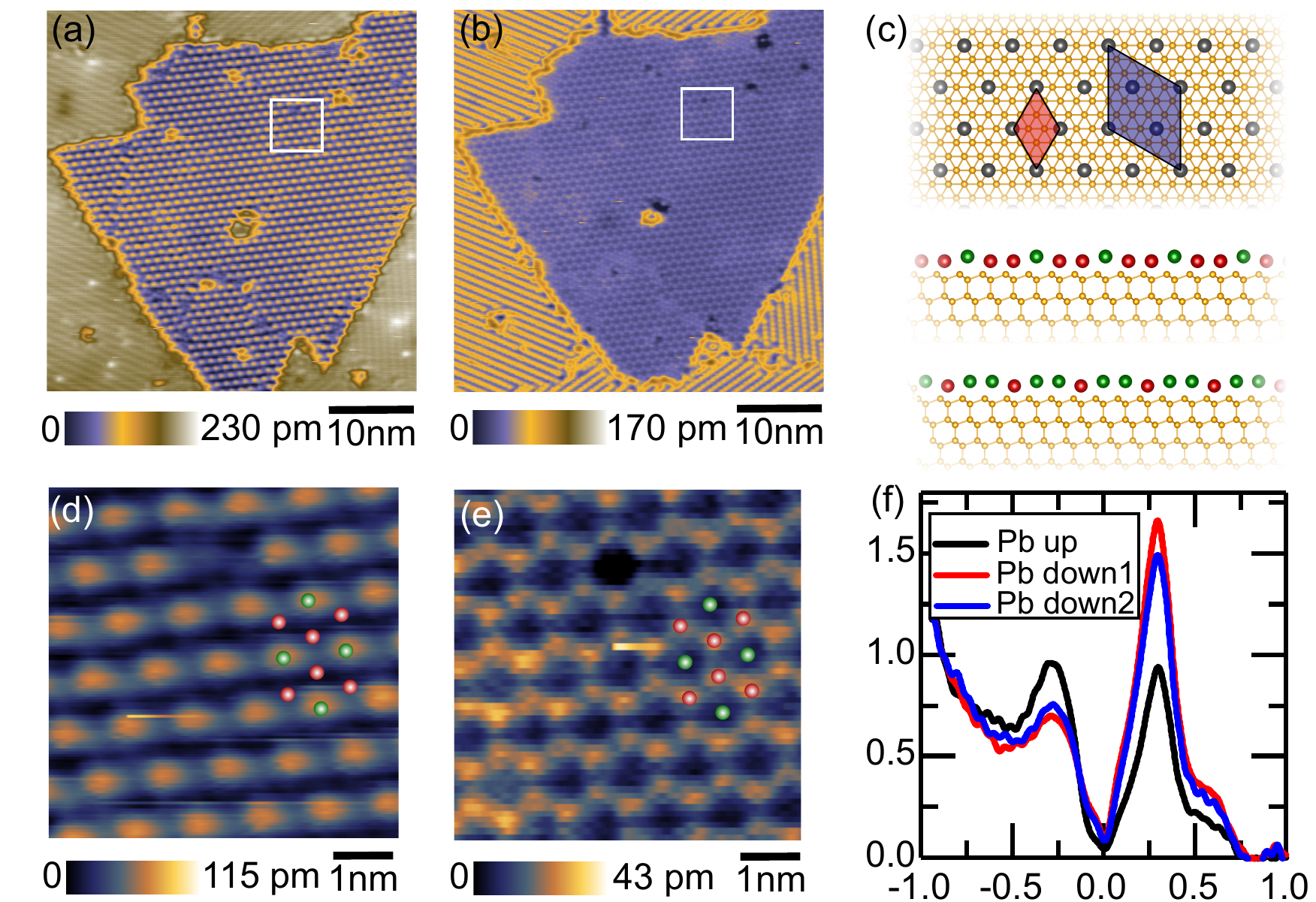}
\caption{(color online) All measurements presented in this figure were carried out at $T= 4.2~$K. (a) STM topography of a $3\times3$ domain measured with $V_{bias}=-1.0~$V and  $I=20~$pA. It shows the sample occupied states. (b) Same domain as in panel (a) but measured with $V_{bias}=+1.0~$V and  $I=20~$pA. It shows the sample unoccupied states. (c) Top: Top view of the high-temperature $\sqrt{3}\times\sqrt{3}$-R30$^\circ$ (red) and low-temperature $3 \times 3$ (blue) unit cells of $\alpha$-Pb/Si(111). Bottom: Side view of the $3 \times 3$ $1u2d$ and $2u1d$ phases. (d) Close-up of the area delimited by the white rectangle in panel (a). It features a triangular lattice of dots associated to the Pb up atoms. The colored balls indicate the theoretical location of the Pb atoms associated with the $1u2d$ phase. (e) same as in (d) but for $V_{bias}=+1.0~$V and  $I=20~$pA. A characteristic honeycomb lattice is seen associated with the two down atoms. (f) $dI/dV$ spectra measured by STS proportionnal to the LDOS of the three different Pb atoms present in the $3\times3$ unit cell denoted Pb-up (black curve), Pb-down1 (red curve) and Pb-down2 (blue curve). Set-point for STS $V_{bias}=-1.0~$V and $I=100~$pA.}\label{topos_point_STS}
\end{figure*}

For a single-band Hubbard model in two-dimensions (2D), depending on the precise lattice geometry and parameters of the hopping integrals, on-site Coulomb repulsion, as well as possible non-local repulsion terms, various phases are theoretically predicted as ground states. These range from paramagnetic metals with or without charge ordering, to insulating states with various magnetic ordering or spin-liquid phase \cite{Balents2010}. 

Such a paradigmatic situation seems at first glance to be realized in 2D in a class of surface crystals, called $\alpha$-phase, consisting in a low density single-layer of tetravalent metal atoms (Pb or Sn) grown on tetravalent semiconducting substrates like Si(111) or Ge(111) \cite{Carpinelli1996,Santoro,PhysRevLett.98.086401,Hansmann2013a}. The atoms are organized in a triangular lattice forming a $\sqrt{3}\times\sqrt{3}$-$R30^{\circ}$ reconstruction in the high-temperature phase. Each metal atom leaves a free electron at T4 sites leading to a single half-filled electronic band confined at the surface and isolated from bulk bands. 
The $\alpha$-phase compounds offer a nice chemical simplification since there are only two different atoms in the unit cell having $\emph{s-p}$ hybridized orbitals. These materials 
are often considered as a prototypical realization of the single band Hubbard model, despite the fact that this aspect has been recently questioned\cite{Tresca2018,Tresca2019}.

The theoretical consideration of the explicit coupling between the electronic, lattice and substrate degrees of freedom seems mandatory for such surface-confined 2D materials, as already established for denser systems\cite{Brun2017}. It is nevertheless at present taken into account only by \emph{ab initio} density functional theory (DFT) methods \cite{PhysRevB.62.1556,PhysRevLett.98.086401,Tresca2018,Tresca2019,PhysRevLett.121.026401,Tresca2019.NbS2} and is neglected in many-body ones (see for instance \cite{Li2011,Hansmann2013b}). Which methods among advanced DFT methods, like DFT+U \cite{PhysRevB.71.035105}, HSE\cite{doi:10.1063/1.1564060,doi:10.1063/1.2204597}, or many-body ones are capable or not of describing the structural, electronic and magnetic ordering taking place in these compounds, should be looked at in details by a precise comparison with all available experimental results for each particular material. 

In a correlated surface material, both electron-phonon (\emph{e-ph}) and electron-electron (\emph{e-e}) effects can contribute to a lattice re-ordering. Experimentally a $3\times3$ reconstruction is observed at low temperature for all $\alpha$-phase compounds except Sn/Si(111). The lattice reconstruction can be inferred from surface x-ray diffraction (SXRD) only when the $\alpha$-phase can be macroscopically and homogeneously grown, as done for Pb/Ge(111) and Sn/Ge(111) \cite{Mascaraque1999,Zhang1999,Bunk1999}. 

The situation is thus more challenging in Pb/Si(111) where a $3\times3$ reconstruction is observed at low temperature by STM \cite{Custance2001b,Brihuega2005,Tresca2018,Adler2019} but where the small size of the $3\times3$ domains coexisting with large $\sqrt7 \times \sqrt3$ regions prevents SXRD or XPS to be used to deduce the refined $3\times3$ structure  \cite{Brihuega2005,Brihuega2007}. In this case the atomic lattice reconstruction and the associated electronic charge ordering has to be solely deduced from STM/STS or AFM experiments which necessitates an error-free protocol. With respect to this issue, the recent studies undertaken in Pb/Si(111) by STM/STS are illuminating as they present conflicting results and different interpretations \cite{Tresca2018,Adler2019}. 

In Ref.\onlinecite{Tresca2018}, Tresca \emph{et al.} showed that at 300~mK the Pb/Si(111)-$3\times$3 ground state is a correlated metal presenting a charge order. 
This correlated metallic state is characterized by a vertical lattice distorsion, having one Pb atom up and two Pb atoms down (called one up-two down $1u2d$ in the following for simplicity). This structural deformation is accompained by a local electronic ordering 
due to the fact that in the $3\times$3 unit cell two electrons among three localize on the Pb up atom while the third 
unpaired electron is shared among the two down Pb atoms ensuring metallicity. 
Note that the same 
$1u2d$ atomic ordering was deduced earlier by Brihuega {\it et al.} from STM topography measurements confronted to DFT calculations in the GGA approximation \cite {Brihuega2005}. 
Additionally, SXRD measurements in Pb/Ge(111) and Sn/Ge(111) showed the same atomic ordering \cite{Zhang1999,Bunk1999}. Surprisingly, the recent STM/STS study by Adler \emph{et al.} of the local charge order in Pb/Si(111) came to the opposite conclusion, promoting a two-up one-down ($2u1d$) atomic ordering \cite{Adler2019}. 

In the present work, we solve the controversy by demonstrating that it is related to an incorrect method to extract information on local order via the lock-in technique.


The described procedure and our considerations are fully general. They should be applicable to any 
material presenting a local atom-dependent charge ordering measured using the STM/STS technique, i.e. oxydes like manganites\cite{Salamon2001} or iridates\cite{Cao2018}, quasi-1D or quasi-2D organic conductors\cite{Wosnitza2007,bookorganics}, Mott insulators\cite{Janod2015}, ultrathin films or monolayers of transition metal dichalcogenides \cite{PhysRevLett.121.026401,Tresca2019.NbS2}, twisted graphene bilayers \cite{bilayergraphene} and so on.

\section{Experimental results}

The $\alpha$-Pb/Si(111) single-layer phase was prepared using a well-known procedure as described in Ref.\onlinecite{Tresca2018}. This procedure results in the formation of a mixed phase consisting of dense, metallic and large $\sqrt7 \times \sqrt3$ domains having a Pb coverage of 1.2 ML adjoining small $\sqrt3 \times \sqrt3$ regions of size less than 100~nm having a Pb coverage of 1/3 ML. Below about 86~K, the $\sqrt3 \times \sqrt3$ regions transit to a $3 \times 3$ structure \cite{Brihuega2005}. 
Our measurements were performed on $3 \times 3$ domains having a lateral size larger than 25~nm at $T=0.3$~K, 2~K or 4.2~K with metallic PtIr or W tips. For such range of domain sizes we have found that the spectral characteristics presented hereafter are well established. The $I(V)$ spectra were measured far enough from boundaries with neighboring $\sqrt7 \times \sqrt3$ domains in order to probe intrinsic electronic properties. The $dI/dV(V)$ spectra were obtained by numerical derivation of the raw $I(V)$ curves. Negative (positive) bias voltage corresponds to occupied (empty) sample states.

Fig. \ref{topos_point_STS} shows a characteristic $3 \times 3$ domain measured by STM at $T=4.2~$K. We used the same tunneling current ($I=20~$pA) but two different bias voltages ($V_{bias}=-1.0~$V in panel (a) and $V_{bias}=+1.0~$V in panel (b)). Strikingly, a contrast reversal exists between panels (a) and (b). In the panel (a) the maxima of the $z(x,y)$ signal of the STM constant current topography features a triangular lattice of bright dots while in panel (b) it forms a honeycomb lattice. Moreover, the location of the bright dots in panel (a) corresponds to the location of the darker dots in panel (b). This can be seen better by inspecting the panels (d) and (e) presenting a close-up of the region situated in the white square. The point defect present in this region features the absence of a bright dot in panel (d) and a darker dot in panel (e). It is presumably a Pb vacancy. Going away from this point defect enables a site-by-site comparison between the panels (d) and (e). 

Our detailed analysis of the experimental results, supplemented by state-of-the-art {\it ab initio} DFT calculations presented below, allows us to conclude that the bright dots in panel (a) and (d) are associated to the Pb-up sites while the two bright dots forming the honeycomb lattice in panel (b) and (e) correspond to the atomic sites of the two Pb-down atoms. For clarity, green and red dots are superposed on STM topographies (d) and (e) to represent the expected locations of the up and down atoms respectively. Nevertheless as we show now, this conclusion cannot solely be deduced from STM measurements. Indeed the characteristic variation in height of the $z(x,y)$ signal measured in panel (d) (respectively (e)) is very small: it is only of about $0.20~$\AA. This small value suggests that there could be in principle two effects contributing to it. There could be i) a difference in the vertical height (i.e. in the direction perpendicular to the surface) of the three Pb atoms inside the $3 \times 3$ unit cell and/or ii) a difference in their integrated LDOS for the occupied (resp. unoccupied) states. As we show hereafter, only the detailed comparison with advanced {\it ab initio} DFT calculations enables us to conclude which effects take place and whether there is or not an atomic reconstruction. Nevertheless, since we aim at concluding whether there is or not an atomic reconstruction accompanied by possible charge ordering on the various Pb sites, we need to perform site-dependent STS measurements in order to probe the corresponding spatial variations of the LDOS at the atomic scale.

Site-dependent STS measurements are presented in panel (f) of Fig.~\ref{topos_point_STS}. The characteristic $dI/dV$ spectra of the three different Pb sites seen by STM inside the $3 \times 3$ unit cell are shown. These spectra are consistent with the spatially averaged $dI/dV$ spectrum presented in our previous work (see Fig.~1c in Ref.\onlinecite{Tresca2018}). In particular all three spectra reveal a strongly depressed DOS at $\varepsilon_F$ not reaching zero, which is a hallmark of the correlated semimetallic character of the surface. This depressed DOS at $\varepsilon_F$ is surrounded by two prominent peaks located almost symmetrically with respect to $\varepsilon_F$ at $-0.3~$eV and $+0.3~$eV. At larger binding energy, i.e. between $-1.0<V<-0.5$~eV, the conductance strongly increases due to the LDOS associated to Si bands \cite{Tresca2018}. In the unoccupied states a sharp drop in conductance occurs above the main peak and negative differential conductance is measured between $\simeq 750$~meV and 1~V. 

\subsection{Probing charge ordering at the atomic scale through LDOS maps with scanning tunneling spectroscopy: methodological considerations}\label{methods}

Since we need and want to compare with one another the LDOS measured on each of the 3 different Pb sites seen in the $3 \times 3$ unit cell by STM, we need to be sure that the site-dependent $dI/dV$ spectra are measured in a way that enables such a spatial comparison. Ideally, according to the Tersoff-Hamann theory of STM, the LDOS should be probed at constant tip-sample height to enable such a correct spatial comparison \cite{TersoffHamann}. If this condition is fulfilled, then each $dI/dV(V)$ spectrum measured over the surface is proportionnal to the sample LDOS and a direct spatial comparison is possible. Let us note that the contribution of the tip DOS to the $dI/dV(V)$ signal is neglected in the Tersoff-Hamman approximation: the tip DOS is assumed to be constant around $\varepsilon_F$. Furthermore, it is also assumed that the spectroscopic measurements are performed in a small enough energy range around $\varepsilon_F$ such that the transmission coefficient does not vary appreciably in this energy range. This is typically verified for an energy interval of [-1;+1]~eV around the Fermi level and a tip work function larger than $4$~eV.

In practice, performing constant-height $I(V)$ spectroscopies with STM is challenging. The reason is that this condition requires to determine precisely a plane parallel to the sample surface along which the tip can be scanned with open feedback loop. Due to the small thermal drift existing along the $z(x,y)$ direction it is usually not possible with this method to perform more than several hundreds of $I(V,x,y)$ spectra. Thus full two-dimensionnal $I(V,x,y)$ spectroscopy grids with high spatial resolution over areas larger than few nm$^2$ are usually impossible. It is nevertheless possible to perform STS measurements under constant tip height using a lock-in technique for a given voltage $V$. This is much less time-consuming and has been demonstrated for a variety of systems. This issue was first discussed in the context of surface states of noble metals \cite{Li1997,Ziegler2009}. The LDOS above adatoms \cite{Ziegler2009} and single molecules \cite{Lu2003} over small areas could also be measured this way. An additional method going beyond constant-height measurements has also been proposed to deal with the case of nano-objects adsorbed on surfaces and improve the deficiency of the constant-height method in this case \cite{Reecht2017}.

The principle of lock-in the detection technique is to acquire directly the $dI/dV(V,x,y)$ signal by adding a small $ac$ voltage $V_{ac}cos(\omega t)$ to the DC bias voltage $V$ applied between tip and sample. The $dI/dV(V,x,y)$ signal is obtained from a demodulation by the lock-in amplifier according to the first-order approximate relation: \\ $I(V+V_{ac},x,y)$ = $I(V,x,y) + dI/dV(V,x,y) V_{ac}cos(\omega t)$. This technique can be used in constant height mode with open feedback loop as mentionned above. In this case it leads to correct LDOS measurements since it fulfils the Tersoff-Hamman condition. Nevertheless, stringent time limitations exist for the STS measurements in a dense grid mode due to the open feedback loop, as explained above. In contrast, if the lock-in technique is used in closed feedback loop while scanning the STM tip in constant current mode, it fails to reveal the true LDOS for energy-dependent and/or site-dependent LDOS, as is usually the case for materials presenting (strong) electronic correlations together with charge ordering. A very good example of this failure is provided by the data of Adler et al. presented in Ref.~\cite{Adler2019} and reproduced in the present figure~Fig.~\ref{comp_dI_dV} for our correlated 2D Pb/Si(111) material. The reason of this failure is that the constant current condition dramatically changes and even can reverse the relative LDOS weight of atomic sites presenting different LDOS energy dependences and/or charge ordering. \\

If one aims at performing a dense two-dimensional and full $I(V,x,y)$ spectroscopy grid, as we need here, that typically contains from several thousands to several tens of thousands of individual $I(V,x,y)$ spectra (taking from about 10 to several tens of hours of measurement), the usual trick is to work under closed feedback loop conditions by choosing an appropriate common set-point $(I_0,V_0)$ for each $I(V,x,y)$ spectroscopy. This last method has been widely used in STS experiment since about twenty years, including for instance complex materials where electronic correlations play an important role such as high-temperature cuprate superconductors \cite{Fischer}. In this case the $dI/dV(V,x,y)$ spectra are obtained by numerical derivation of the $I(V,x,y)$ curves. It can directly be seen that for this latter method to be correct, the choice of the $(I_0,V_0)$ set-point should qualitatively reflect the constant-height condition. Thus in each particular situation and material, depending on the energy range that is to be measured, care should be taken regarding the choice of the $(I_0,V_0)$ set-point. Fortunately, for many materials there often exists an energy range $[eV_1,eV_2]$ for which the LDOS is homogeneous over the area of interest, which enables a post-normalization procedure of the $dI/dV(V,x,y)$ spectra. This corresponds to the procedure that we have followed and that will be further explained below.

\subsection{Probing charge ordering at the atomic scale through LDOS maps with scanning tunneling spectroscopy: the case of 1/3 monolayer Pb/Si(111)}\label{dI_dV_maps}

\begin{figure*}
\centering 
{\includegraphics[width=0.85\textwidth]{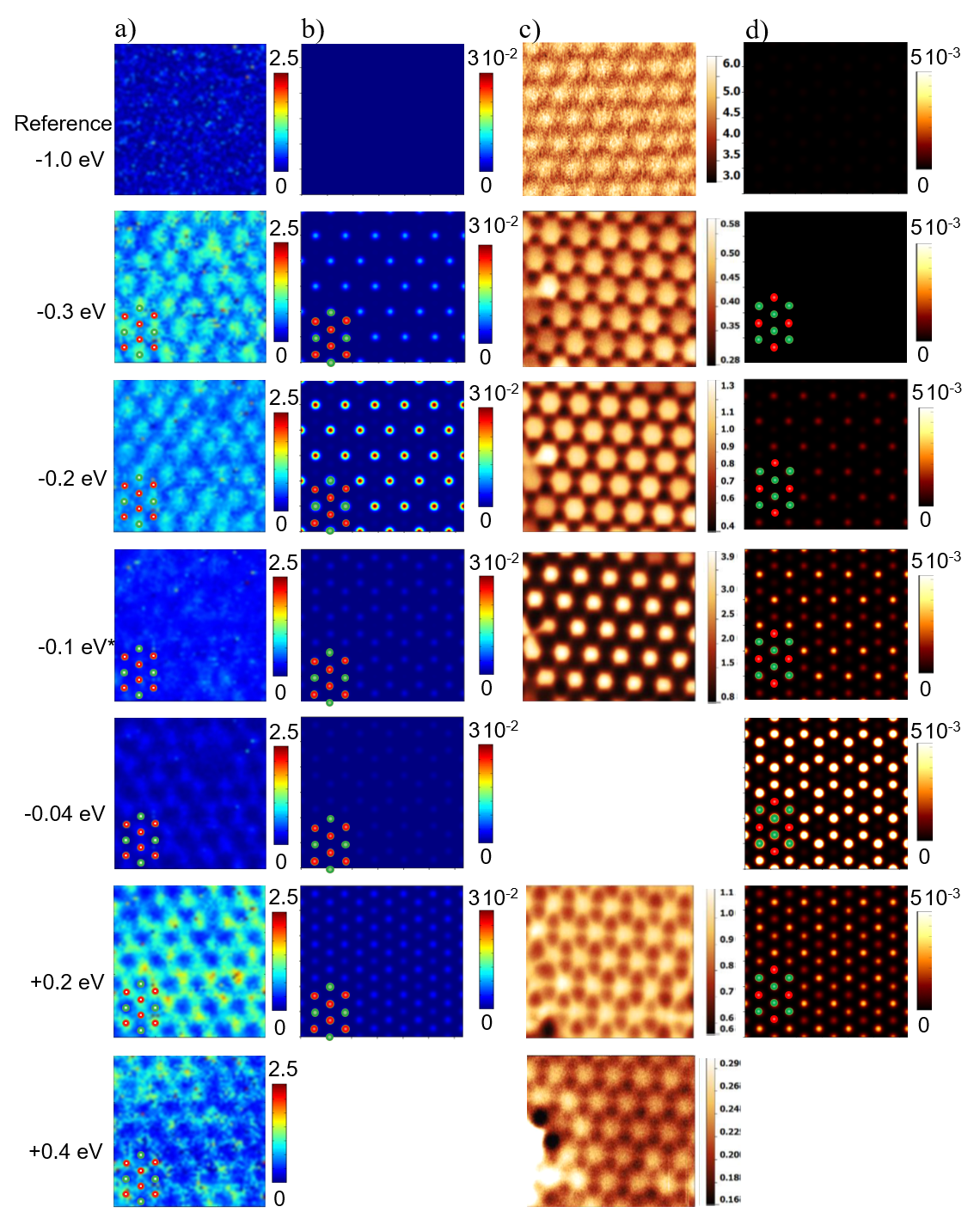}}
\caption{(color online) Comparison of the energy-resolved LDOS maps of the $3\times3$-Pb/Si(111) phase measured at $4$~K by STS using two different methods, confronted to advanced DFT calculations for the optimized $1u2d$ 
and $2u1d$ 
phases. Column a) represents this work.  It shows $dI/dV(E,x,y)$ maps numerically derivated from a grid of $200\times200$ $I(V,x,y)$ curves acquired between [-1;+1]~eV with a common set-point for each spectroscopy at $I_0=100$~pA, $V_0=-1.0$~V. Each row has a different bias adopted for the measurement, the corresponding energy $E$ being indicated in eV. The normalized conductance color scale is identical for all these panels and is shown by a vertical bar (having $dI/dV(E,x,y)$ values between 0 and 2.5) on the right of each map. Column b) is the theoretical LDOS for the $1u2d$ phase while in column d) the theoretical LDOS for the "proposed" $2u1d$ system is reported. The theoretical panels are presented with a common color scale for each phase, namely between 0 and 3~$10^{-2}$ states/Ry for the $1u2d$ (respectively 0 and 5~$10^{-3}$ states/Ry for the $2u1d$). The colored balls in columns a) and b) feature the theoretical location of the Pb atoms associated to the ``one up-two down'' phase. In column c) we reproduce the $dI/dV(E,x,y)$ maps by Adler {\it et al.}\cite{Adler2019}, for comparison. The measurements of by Adler {\it et al.}are obtained by using a lock-in technique and acquired in closed feedback loop together with the STM topography at the corresponding voltage. In these measurements the conductance color scale vary for each energy $E$, which we indicated accordingly. Note also that these measurements are rotated by almost 90$^{\circ}$ with respect to ours.\\ *) For the energy $E=-0.1~$eV, the map in column a) was in fact obtained at an exact energy of $-0.115~$eV.  
}\label{comp_dI_dV}
\end{figure*}

From Fig.~\ref{topos_point_STS} we see that the actual measured bandwidth of the $3\times3$-Pb/Si(111) phase is about [-0.5;+0.5]~eV. In order to extract a meaningful local $dI/dV(E,x,y)$ signal proportionnal the intra-unit cell LDOS, we have chosen to measure a $200\times200$ grid of single $I(V,x,y)$ curves between [-1;+1]~eV with a common set-point for each STS spectroscopy at $I_0=100$~pA, $V_0=-1.0$~V. This dense set of $dI/dV(E,x,y)$ data is obtained from numerical derivation of the corresponding single $I(V,x,y)$ curves. The set-point in voltage is chosen far in energy from the 2D surface band of interest. In particular, as our DFT calculations presented below show, this set-point is chosen well enough in the silicon bulk valence band so that no peculiar site-dependence exists in the LDOS for $E \approx eV_0$. This means that it is possible to post-normalize the measured $dI/dV(V,x,y)$ spectra at the energy $E \approx eV_0$. This is exactly the procedure that we have followed here. One can also notice that the same normalization procedure was performed in Fig.~\ref{topos_point_STS}f). As a consequence, our reference $dI/dV(E=-1eV,x,y)$ map shown in the first row of Fig.~\ref{comp_dI_dV} in panel a) presents an homogeneous aspect. This spectroscopic behavior is well reproduced by our optimized DFT calculations presented in panel b), evaluating the LDOS for the $1u2d$ phase at a height of 3~\AA~from the topmost Pb atoms.

The panels in column a) 
of the figure~\ref{comp_dI_dV} present $dI/dV(E,x,y)$ maps obtained following this protocol for various energy $E$ running through the whole 2D correlated surface bands. Let us note that the conductance color scale of each $dI/dV(E,x,y)$ map is identical (having conductance values between 0 and 2.5). This enables a proper quantitative comparison of the LDOS measured at different energies. The three different Pb atoms are superposed with lighter (green) color for the up atoms and darker (red) color for the down ones. These experimental $dI/dV(E,x,y)$ maps can be directly compared to the panels in column b) 
showing our optimized DFT calculations of the LDOS for the $1u2d$ phase evaluated at a height of 3~\AA~from the topmost Pb atoms.

At larger binding energy $E\approx-0.3$~eV, close to the peak observed in the LDOS of the occupied states, 
our measurements show that the LDOS is maximum on the up atoms, in agreement with the panel f) of the Fig.~\ref{topos_point_STS}. For this energy, the dominant LDOS signal forms a triangular lattice of bright dots associated to the Pb up atoms. This LDOS behavior is precisely reproduced by DFT calculations of the $1u2d$ phase, 
even if theoretically, the LDOS maximum on the Pb up atoms is obtained for slightly smaller energy $E\approx-0.2$~eV. 
When the binding energy $E$ reduces for $-0.3\le E \le -0.115$~eV, the experimental LDOS on all sites reduces. Simultaneously, the relative LDOS weight between the up and down atoms also reduces. When one reaches $E \approx -0.115$~eV 
the LDOS becomes approximately equal on all three Pb sites, as seen by the homogeneous color, in agreement with Fig.~\ref{topos_point_STS}f. Theoretically, such a behavior is also predicted to occur around $E = -0.1$~eV as shown by our theoretical predictions. 

For $E \ge -0.115$~eV, the relative LDOS weight reverses, the LDOS becoming larger on the down atoms than on the up ones, as expected from Fig.~\ref{topos_point_STS}f). 
This behavior is well reproduced by DFT calculations. 
Between $0\le E \le 0.3$~eV the LDOS rises on all three Pb sites and the relative weight between the down and up sites also increases. This is illustrated for $E \approx +0.2$~eV 
where one sees that the LDOS weight on the two down atoms is almost equal and dominates. As a result, the dominant LDOS signal forms a honeycomb lattice. This honeycomb lattice presents a complete site-reversed contrast with respect to the triangular lattice seen 
for $E=-0.3$~eV. The theoretical results 
agree with this picture. When reaching energies close to the upper band edge, both the LDOS and relative weight between the up and down sites diminish. 
Furthermore, slight inhomogeneities between sites of the same type become visible, probably induced by disorder effects. The theoretical bandwidth being slightly smaller than the measured one, the comparison cannot be made for this latter energy.

\subsection{Probing charge ordering at the atomic scale through LDOS maps with scanning tunneling spectroscopy: on the misuse of the lock-in technique}

Our atomically resolved LDOS measurements inside the Pb-$3\times3$ unit cell are in strong constrast with the previously reported STS results by Adler {\it et al.}~\cite{Adler2019}. A detailed comparison is provided for similar energies in Fig.~\ref{comp_dI_dV}. For large binding energies ($E\approx-0.3$~eV) the LDOS map 
is similar to ours, revealing a triangular lattice of bright dots. However, as the energy increases the LDOS contrast never qualitatively changes in Adler {\it et al.}'s data. 
For instance, the LDOS weight reduction around $\varepsilon_F$ is not observed, nor the LDOS reversal between the up and down atoms for $E \ge -0.115$~eV. Instead, Adler {\it et al.}'s measurements suggest that the LDOS is always dominated by the same type of atoms for all energies inside the bandwidth. This is physically impossible. This is also opposite to our experimental measurements and theoretical analysis, the latter being detailed below in Sec.\ref{sec_th}, since in our modelling we also considered the possibility of the formation of the $2u1d$ phase. 

Furthermore, the $dI/dV(E,x,y)$ maps presented by Adler {\it et al.} are themselves in contradiction with their own site-dependent $dI/dV(V,x,y)$ spectra presented in the figure~S4 of the supplementary materials of Ref.~\cite{Adler2019}. Indeed from this figure~S4 the site-dependent $dI/dV(V,x,y)$ spectra show a qualitative behavior similar to the one presented in Fig.~\ref{topos_point_STS}f). This shows that for $E \lessapprox -0.1$~eV their $dI/dV(E,x,y)$ maps should be dominated by the LDOS measured on what they called ``center'' atoms (Pb up atoms in our case). Then an almost equal LDOS weight should be observed around $\approx-0.1$~eV between the denoted ``center'' and ``corner'' atoms (Pb down atoms in our case), followed by an LDOS contrast reversal for $\approx-0.1 \le E \le \approx+0.3$~eV. This expected energy dependence is clearly not what is measured in the $dI/dV(E,x,y)$ maps of Ref.\onlinecite{Adler2019}, 
that we reproduced in Fig.~\ref{comp_dI_dV}. This experimental artefact, leading to wrong LDOS measurements, is due the improper use of the lock-in technique in closed feedback loop while acquiring the STM topography for a correlated 2D material presenting site-dependent LDOS and charge ordering, as we explained before in section~\ref{methods}.

Another unfortunate consequence of the misuse of the lock-in technique to acquire $dI/dV(V,x,y)$ maps in closed feedback loop deals with quasi-particle interference measurements. Quasi-particle interference (QPI) measurements are a phase-coherent manifestation of the scattering of elementary electron- or hole-excitations by defects\cite{SimonFTSTS}. These QPI measurements are typically obtained by performing the Fourier transform of a $dI/dV(V,x,y)$ map measured for a given energy. 
Since we just saw that $dI/dV(V,x,y)$ maps acquired in closed feedback loop are incorrect for site-dependent LDOS, one deduces that QPI maps will also be wrong. This is particularly important to be noticed since this technique is broadly used to infer information about order parameter symmetry, topology, spin and band-structure properties \cite{Chen2017,Lin2020}. We thus may consider with care the recent QPI experiments done this way in $\alpha$-phase materials\cite{Adler2019,Ming}. Evidence for this situation is provided by comparing the QPI maps at the Fermi level for Pb-$3\times3$/Si(111) obtained using our method (see Fig.~3 in reference~\cite{Tresca2018}) and Adler's one (see Fig.~3 in reference~\cite{Adler2019}).

In order to support the presented experimental STM/STS results and their interpretation, we now present and discuss in details the theoretical investigations of the $1/3$~ML $\alpha$-Pb/Si(111) phase using advanced DFT calculations.

\section{Theory: modelling the structural and electronic properties of the correlated 1/3 monolayer Pb/Si(111)}\label{sec_th}

\subsection{Computational details}
As in Ref.\onlinecite{Tresca2019} we model the Pb/Si(111) surfaces by
considering a layer of Pb atoms on top of 3 or 6 Si bilayers. The bottom dangling bonds are capped with hydrogen atoms fixed to the relaxed positions obtained by capping one side of the pristine Si surface.
The atomic position of the first five (four) Si-substrate layers
below the Pb single layer are optimized whereas the remaining seven (two) layers are fixed to the Si bulk positions. More than 15~\AA~of
vacuum are included.

Density functional theory (DFT) calculations are performed with \textsc{Quantum-Espresso}\cite{QEcode, QE-2017} and \textsc{Crystal17}\cite{doi:10.1002/wcms.1360,cryman}
 codes. For plane wave calculations, we used ultrasoft pseudopotentials with the same settings of Ref.\onlinecite{Tresca2018}. We used the following semi-local approximations for the exchange and correlation kernel: the local density approximation (LDA), the generalized gradient approximation (GGA) and the GGA+U approximation with an energy cutoff up to 45~Ry. Integration over Brillouin zone (BZ) was performed using uniform $10(6)\times 10(6)\times 1$ Monkhorst and Pack grids\cite{PhysRevB.13.5188} for the $\sqrt{3}\times\sqrt{3}$-R30$^o$(3$\times$3) and a 0.001~Ry Gaussian smearing.

Hybrid-functionals calculations in plane waves for cells as large as those considered here are hardly feasible. Thus, HSE06\cite{doi:10.1063/1.1564060,doi:10.1063/1.2204597} (non-relativistic) calculations were performed by using the \textsc{Crystal17}~\cite{doi:10.1002/wcms.1360,cryman} code with Gaussian basis sets exactly as in Ref.\onlinecite{Tresca2019}. The basis sets used for these calculations have been directly downloaded from the \textsc{Crystal} web site. For Pb we used pseudopotentials\cite{Sophia2013,PhysRevB.74.073101} from the \textsc{Crystal} distribution, while for Si and the capping H an all-electron m-6-311G(d)\cite{doi:10.1063/1.2085170,Pernot2015} and TZVP\cite{Peintinger2012} have been respectively adopted. 

Integration over BZ was performed with the same k-mesh density as in the plane wave calculation and a Fermi-Dirac smearing of 0.0005~Ha. The integration threshold was set to $10^{-7}$ for integrals in the Coulomb series and $10^{-7}$, $10^{-15}$ and $10^{-30}$ for the exchange ones (see Ref.\onlinecite{doi:10.1002/wcms.1360,cryman} for more details).

In this framework, we optimize the internal coordinates. 

Relativistic effects are not implemented in the \textsc{Crystal17} code. To overcome
this difficulty, in  analogy with Ref.\onlinecite{Tresca2019}, we fit the non-relativistic 
HSE06 electronic structure and Fermi surfaces at HSE06 fixed geometry in a DFT+U formalism with $U$ on the Pb and Si $p$ states, and then apply non-collinear spin-orbit on top at fixed atomic coordinates  in a DFT+U+SOC calculation. This is possible because relativistic effects are negligible in the atomic relaxation process\cite{Tresca2018,Tresca2019}, vice versa a good description of the energy gap of Si is necessary for the structural prediction. 

\subsection{Determination of the ground state}

As known, at low temperature the $\alpha$-Pb/Si(111) shows a charge density wave transition (CDW) between the $\sqrt{3}\times\sqrt{3}$-R30$^o$ periodicity to the $3\times 3$ ordering with three inequivalent Pb atoms\cite{Carpinelli1996,PhysRevB.57.14758,MASCARAQUE1999337,PhysRevLett.94.046101,PhysRevB.75.155411,Tejeda_2007,Cudazzo2008747,PhysRevLett.82.2524,Tresca2018,Adler2019}. For years the $3\times 3$ phase in Pb/Si(111) has been interpreted as a deformation in which one Pb atom is higher and the other two closer to the substrate, in analogy with the similar Pb/Ge(111)\cite{Carpinelli1996,PhysRevB.57.14758,PhysRevLett.82.2524,Tejeda_2007,Tresca2019}. This reconstruction is usually labeled $1$ up and $2$ down ($1u2d$). Our previous work showed that not only electron-phonon coupling drives this reconstruction but also on-site electron-electron repulsion \cite{Tresca2018}. As mentioned, this aspect has been recently questioned 
by an experimental/theoretical work claiming 
that the low temperature Pb/Si(111) ground state would appear to be the $2$ up and $1$ down structure ($2u1d$)\cite{Adler2019}.

In order to reconcile this disagreement, as first step, we studied the stability of both $1u2d$ and $2u1d$ CDW with different functionals and different substrate thickness modelization. Results are summarized in Tab.\ref{tabth}.
\begin{table*}[]
\footnotesize
\begin{center}
\begin{tabular}{|l|c|c|c||c|c|c|}
\multicolumn{1}{l}{ } &\multicolumn{3}{c}{3 Bi-layers} & \multicolumn{3}{c}{6 Bi-layers}\\

\hline
        & $\sqrt{3}\times\sqrt{3}$  & $3\times3$ (1u2d) & $3\times3$ (2u1d) & $\sqrt{3}\times\sqrt{3}$  & $3\times3$ (1u2d) & $3\times3$ (2u1d) \\
\hline
LDA     &   {\bf 0.000}\cite{} &  +0.000 & +0.000 & {\bf 0.000}   &  +0.000   &   +0.001  \\
\hline
GGA     &    +0.005  &  {\bf 0.000} & +0.004   & +0.001   & {\bf 0.000}    & +0.001 \\
\hline
GGA+U (U=2~eV)  &  +0.008  &  {\bf 0.000} & +0.007 & +0.003 & {\bf 0.000}    &  +0.003  \\
\hline
GGA+U (U=8~eV)   & +0.296   & +0.141  & {\bf 0.000} & +0.155  & +0.137    & {\bf 0.000}  \\ 
\hline
HSE06   &                 +0.026    &  {\bf 0.000}   & $NS$  & +0.023 &  {\bf 0.000}   & $NS$   \\
\hline
\end{tabular}
\end{center}
\caption{Stability of the different phases of 1/3 monolayer Pb/Si(111) simulated with 3 and 6 Si-bilayers. The high temperature phase is labeled $\sqrt{3}\times\sqrt{3}$. The energy differences are referred with respect to the ground state (in bold). All energies are expressed in eV/Pb. The acronym $NS$ means ”not stable”.
}
\label{tabth}
\end{table*}

At the semilocal functional level (LDA and GGA), for the $1u2d$ CDW phase, we obtain the results already known in literature\cite{Cudazzo2008747,PhysRevB.94.224418,Tresca2018} and we find that the $2u1d$ phase is not stable. 

We repeated the calculation by including a moderate and strong Hubbard term on Pb sites (2 and 8 eV respectively). As summarized in Tab.\ref{tabth} the only case able to describe the $2u1d$ CDW phase as ground state is the inclusion of a correlation term of 8~eV on Pb atoms. The eventual physical justification for such a high value is far from trivial and makes this hypothesis a mere stylistic exercise. 
Furthermore in such a case the height difference found between "up" and "down" Pb atoms is $\sim 1.9~$\AA, a completely unrealistic value. 

On the contrary, predictions with hybrid functionals (HSE06) have proven to be extremely accurate for the similar compound $\alpha$-Pb/Ge(111)\cite{Tresca2019}, due to the strong hybridization of the Pb valence electronic wavefunctions with the ones of the substrate atoms, as also noticed previously for Sn/Ge(111) and Sn/Si(111)\cite{Lee2013,Lee2014}. In the $\alpha$-Pb/Si(111) case, the HSE06 calculations predict the $1u2d$ CDW to be the ground state. The $2u1d$ phase is never stable. Even starting from the $2u1d$, during the relaxation process, the system escapes from this configuration to adopt the $1u2d$ deformation.

\begin{figure}[h!]
\centering 
\includegraphics[width=\linewidth]{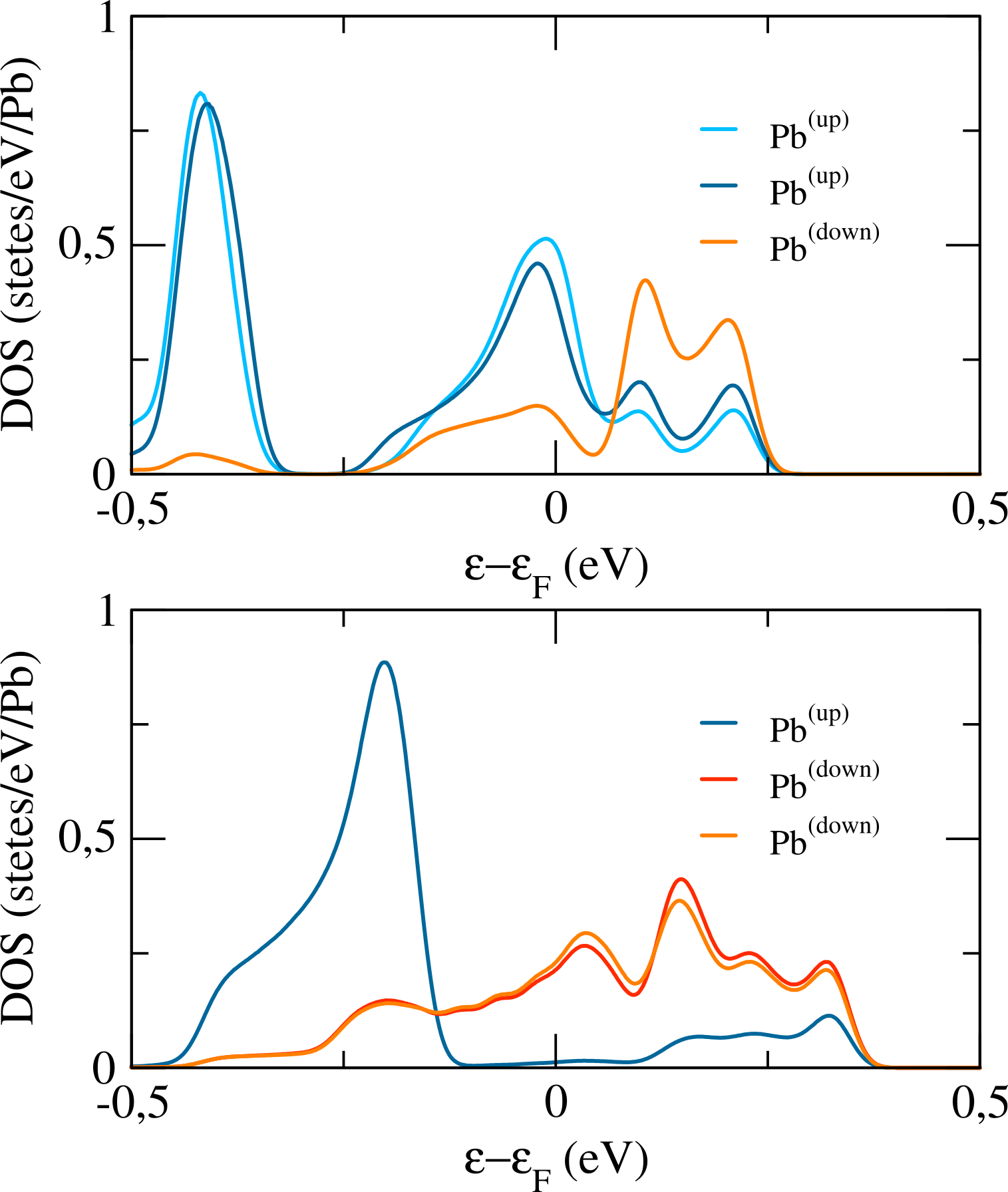}
\caption{Predicted Pb atom-projected DOS for the "ideal" $2u1d$ CDW (top) and for the $1u2d$ one (bottom). In the top panel, the high peak at $\sim-0.42$~eV is an artificial structure due to the impossibility to relax of the system. Blue curves features the Pb$^{up}$ contributions, orange the Pb$^{down}$ ones. Please note that the "balanced point", defined as the energy around which the projected DOS of the up atom(s) cross the one of the down atom(s), is located at opposite energies for the $2u1d$ and $1u2d$ CDW (positive and negative respectively).}\label{PDOS}
\end{figure}


Even assuming an unlikely failure of the well-known predictive power of the DFT, we tried to calculate the electronic properties of the hypothetical $2u1d$ system. We model this material by considering the substrate in its ideal positions. Following the atomic height variations measured by STM in Ref.~\onlinecite{Adler2019}, we assume one Pb atom in the T4 site at a distance of 1.90~\AA~from the top Si atoms while the other two Pb atoms are also located in T4 positions but at a height of 0.67\AA~above the first Pb one.

Comparing our theoretical DOS prediction for the $2u1d$ "proposed" system (see Fig.\ref{PDOS}-top) with the experimental Pb-resolved STS spectra (see Fig.\ref{topos_point_STS}f), it seems obvious that the theoretical Pb contribution should be inverted. In fact, our theoretical prediction for the "ideal" $2u1d$ system shows a double "up" contribution higher in the filled region and lower in the empty states region. The "down" contribution has a single-component and shows opposite intensity behaviour, in clear contrast with experience (see Fig.\ref{topos_point_STS}f and the experimental LDOS behavior presented in Fig.~\ref{comp_dI_dV}). 



Another important aspect discussed already in section~\ref{dI_dV_maps} is the presence of a "balanced point": by this term we mean the energy for which the partial DOS contributions from the different Pb atoms are approximately equal in the unit cell. Therefore, at this energy, the LDOS map should appear (almost) "flat" and homogeneous. We identify this "balanced point" to be  $\sim+0.07$~eV in the "ideal" $2u1d$ structure and $\sim-0.13$~eV for the HSE06-relaxed $1u2d$ one. The "balanced point" is predicted to be positive in $2u1d$ and negative in the $1u2d$ case. This negative value of $\sim-0.13$~eV for the ``balanced point'' of the $1u2d$ phase is in excellent agreement with our experimental results where it occurs around $\sim-0.115$~eV (compare first the spectra in Fig.~\ref{topos_point_STS}f with the ones in Fig.~\ref{PDOS}-bottom, and then look at the homogeneous conductance map seen in Fig.~\ref{comp_dI_dV} column a) for E=-0.1~eV ). 

In order to enable a direct comparison between theoretical atom-projected DOS and experimental LDOS maps, we evaluated the {\it ab initio} theoretical LDOS at a height of 3~\AA~from the topmost Pb atom(s). The results are summarized in Fig.~\ref{comp_dI_dV} in columns b) 
for the $1u2d$ phase and d) 
for the $2u1d$ one.


These results confirm once more the correctness of our measurement method and the fact that low-temperature $\alpha$-Pb/Si(111) adopt the $1u2d$ CDW phase. In fact, the $1u2d$ theoretical results are in good agreement with experiments, apart for the absolute values of the bias at which we make the comparisons. This last aspect is due to a non perfect description of the electronic renormalization, but the general trend is properly described: for low energies (see 
Fig.\ref{comp_dI_dV}) we observe a triangular lattice linked to the filled Pb-up states. Near the "balanced point" the intensity of the LDOS map is small 
and rather homogeneous. For larger energies, the transition to the honeycomb pattern, linked to the Pb-down states, is evident. 

For what concerns the ideal $2u1d$ configuration, the triangular lattice is evident around -0.2~eV. 
However the honeycomb pattern, related to the 2 Pb-up atoms, appears immediately for higher energies and remains extremely evident up to energies very close to the Fermi level. 
Furthermore, a contrast reversal in the LDOS map should also occur in this case due to the existence of a ``balanced point''.

Finally, thanks to our ground state DFT predictions (see Tab.\ref{tabth}) and to the spectroscopic results in good agreement with the DFT ones, we can definitively conclude that the correlated $\alpha$-Pb/Si(111) low temperature CDW phase is the 1 up-2 down reconstruction. 
 
\section{Conclusion}

In this work we addressed general issues and showed correct ways to extract the local charge order in two-dimensional systems from 
STM/STS measurements. We showed that a particularly important care should be taken when the charge order presents spatial variations at the atomic scale inside the unit cell. In such situations we showed that the widely used lock-in technique, performed while acquiring an STM topography in closed feedback loop, fails to extract this local charge order from STS dI/dV differential conductance maps. In contrast a correct method is either to perform a constant height measurement or to perform a full grid of dI/dV(V) spectroscopies, using a setpoint for the bias voltage outside the bandwidth of the two-dimensional material where the 
LDOS is expected to be spatially homogeneous. 

We have demonstrated our ideas using as a paradigmatic example of two-dimensional material the 1/3 single-layer $\sqrt{3}\times\sqrt{3}$-$R30^{\circ}$-Pb/Si(111) phase, which presents in addition strong electron-electron interaction. As large areas of this phase cannot be grown, the charge ordering in this system is not accessible to macroscopic probes like ARPES 
or grazing x-ray diffraction, but can be determined only by local probes. 

Our combined experimental and theoretical study enabled to establish definitely that the low-temperature charge ordering in $\sqrt{3}\times\sqrt{3}$-$R30^{\circ}$-Pb/Si(111) is associated with the one-Pb-up two-Pb-down $3\times3$ atomic reconstruction. The detailed comparison between STS energy-dependent LDOS maps and state-of-the-art DFT calculations agrees very well. Additionally, we show that this structural and charge re-ordering inside the $3\times 3$ unit cell is equally driven by electron-electron interaction and the coupling to the surface. An important output of our results is that theoretical many-body approaches, neglecting the coupling to the substrate degrees of freedom, miss a proper description of these two-dimensional correlated systems located at the surface plane of bulk crystals.

The described procedure and our considerations are fully general. They should be applicable to any 
material presenting a local atom-dependent charge ordering measured using the STM/STS technique, i.e. oxydes like manganites\cite{Salamon2001} or iridates\cite{Cao2018}, quasi-1D or quasi-2D organic conductors\cite{Wosnitza2007,bookorganics}, Mott insulators\cite{Janod2015}, ultrathin films or monolayers of transition metal dichalcogenides \cite{PhysRevLett.121.026401,Tresca2019.NbS2}, twisted graphene bilayers \cite{bilayergraphene} and so on.

\section{Acknowledgements}
We acknowledge CINECA (ISCRA initiative) for computing resources. CT wants to thank the DSFC of University of L'Aquila for providing him with the equipment to work with. This work was supported by French state funds managed by the ANR within the Investissements d'Avenir programme under reference  ANR-11-IDEX-0004-02, and more specifically within the framework of the Cluster of Excellence MATISSE led by Sorbonne Universit\'es and ANR contract RODESIS ANR-16-CE30-0011-01.

\bibliography{bibliography}{}

\begin{thebibliography}{65}
\expandafter\ifx\csname natexlab\endcsname\relax\def\natexlab#1{#1}\fi
\expandafter\ifx\csname bibnamefont\endcsname\relax
  \def\bibnamefont#1{#1}\fi
\expandafter\ifx\csname bibfnamefont\endcsname\relax
  \def\bibfnamefont#1{#1}\fi
\expandafter\ifx\csname citenamefont\endcsname\relax
  \def\citenamefont#1{#1}\fi
\expandafter\ifx\csname url\endcsname\relax
  \def\url#1{\texttt{#1}}\fi
\expandafter\ifx\csname urlprefix\endcsname\relax\def\urlprefix{URL }\fi
\providecommand{\bibinfo}[2]{#2}
\providecommand{\eprint}[2][]{\url{#2}}

\bibitem[{\citenamefont{Fazekas and Tosatti}(1980)}]{Fazekas1980}
\bibinfo{author}{\bibfnamefont{P.}~\bibnamefont{Fazekas}} \bibnamefont{and}
  \bibinfo{author}{\bibfnamefont{E.}~\bibnamefont{Tosatti}},
  \bibinfo{journal}{Physica B$+$C} \textbf{\bibinfo{volume}{99}},
  \bibinfo{pages}{183} (\bibinfo{year}{1980}),
  \urlprefix\url{https://doi.org/10.1016/0378-4363(80)90229-6}.

\bibitem[{\citenamefont{Carpinelli et~al.}(1996)\citenamefont{Carpinelli,
  Weitering, Plummer, and Stumpf}}]{Carpinelli1996}
\bibinfo{author}{\bibfnamefont{J.~M.} \bibnamefont{Carpinelli}},
  \bibinfo{author}{\bibfnamefont{H.~H.} \bibnamefont{Weitering}},
  \bibinfo{author}{\bibfnamefont{E.~W.} \bibnamefont{Plummer}},
  \bibnamefont{and} \bibinfo{author}{\bibfnamefont{R.}~\bibnamefont{Stumpf}},
  \bibinfo{journal}{Nature} \textbf{\bibinfo{volume}{381}},
  \bibinfo{pages}{398} (\bibinfo{year}{1996}).

\bibitem[{\citenamefont{Carpinelli et~al.}(1997)\citenamefont{Carpinelli,
  Weitering, Bartkowiak, Stumpf, and Plummer}}]{Carpinelli1997}
\bibinfo{author}{\bibfnamefont{J.~M.} \bibnamefont{Carpinelli}},
  \bibinfo{author}{\bibfnamefont{H.~H.} \bibnamefont{Weitering}},
  \bibinfo{author}{\bibfnamefont{M.}~\bibnamefont{Bartkowiak}},
  \bibinfo{author}{\bibfnamefont{R.}~\bibnamefont{Stumpf}}, \bibnamefont{and}
  \bibinfo{author}{\bibfnamefont{E.~W.} \bibnamefont{Plummer}},
  \bibinfo{journal}{Phys. Rev. Lett.} \textbf{\bibinfo{volume}{79}},
  \bibinfo{pages}{2859} (\bibinfo{year}{1997}),
  \urlprefix\url{http://link.aps.org/doi/10.1103/PhysRevLett.79.2859}.

\bibitem[{\citenamefont{Ottaviano et~al.}(2000)\citenamefont{Ottaviano,
  Crivellari, Lozzi, and Santucci}}]{Ottaviano2000}
\bibinfo{author}{\bibfnamefont{L.}~\bibnamefont{Ottaviano}},
  \bibinfo{author}{\bibfnamefont{M.}~\bibnamefont{Crivellari}},
  \bibinfo{author}{\bibfnamefont{L.}~\bibnamefont{Lozzi}}, \bibnamefont{and}
  \bibinfo{author}{\bibfnamefont{S.}~\bibnamefont{Santucci}},
  \bibinfo{journal}{Surface Science} \textbf{\bibinfo{volume}{445}},
  \bibinfo{pages}{L41} (\bibinfo{year}{2000}),
  \urlprefix\url{https://doi.org/10.1016/s0039-6028(99)00974-7}.

\bibitem[{\citenamefont{Brihuega
  et~al.}(2005{\natexlab{a}})\citenamefont{Brihuega, Custance, P\'erez, and
  G\'omez-Rodr\'{\i}guez}}]{PhysRevLett.94.046101}
\bibinfo{author}{\bibfnamefont{I.}~\bibnamefont{Brihuega}},
  \bibinfo{author}{\bibfnamefont{O.}~\bibnamefont{Custance}},
  \bibinfo{author}{\bibfnamefont{R.}~\bibnamefont{P\'erez}}, \bibnamefont{and}
  \bibinfo{author}{\bibfnamefont{J.~M.} \bibnamefont{G\'omez-Rodr\'{\i}guez}},
  \bibinfo{journal}{Phys. Rev. Lett.} \textbf{\bibinfo{volume}{94}},
  \bibinfo{pages}{046101} (\bibinfo{year}{2005}{\natexlab{a}}),
  \urlprefix\url{https://link.aps.org/doi/10.1103/PhysRevLett.94.046101}.

\bibitem[{\citenamefont{Ghiringhelli et~al.}(2012)\citenamefont{Ghiringhelli,
  Tacon, Minola, Blanco-Canosa, Mazzoli, Brookes, Luca, Frano, Hawthorn, He
  et~al.}}]{Ghiringhelli2012}
\bibinfo{author}{\bibfnamefont{G.}~\bibnamefont{Ghiringhelli}},
  \bibinfo{author}{\bibfnamefont{M.~L.} \bibnamefont{Tacon}},
  \bibinfo{author}{\bibfnamefont{M.}~\bibnamefont{Minola}},
  \bibinfo{author}{\bibfnamefont{S.}~\bibnamefont{Blanco-Canosa}},
  \bibinfo{author}{\bibfnamefont{C.}~\bibnamefont{Mazzoli}},
  \bibinfo{author}{\bibfnamefont{N.~B.} \bibnamefont{Brookes}},
  \bibinfo{author}{\bibfnamefont{G.~M.~D.} \bibnamefont{Luca}},
  \bibinfo{author}{\bibfnamefont{A.}~\bibnamefont{Frano}},
  \bibinfo{author}{\bibfnamefont{D.~G.} \bibnamefont{Hawthorn}},
  \bibinfo{author}{\bibfnamefont{F.}~\bibnamefont{He}}, \bibnamefont{et~al.},
  \bibinfo{journal}{Science} \textbf{\bibinfo{volume}{337}},
  \bibinfo{pages}{821} (\bibinfo{year}{2012}),
  \urlprefix\url{https://doi.org/10.1126/science.1223532}.

\bibitem[{\citenamefont{Comin et~al.}(2014)\citenamefont{Comin, Frano, Yee,
  Yoshida, Eisaki, Schierle, Weschke, Sutarto, He, Soumyanarayanan
  et~al.}}]{Comin2014}
\bibinfo{author}{\bibfnamefont{R.}~\bibnamefont{Comin}},
  \bibinfo{author}{\bibfnamefont{A.}~\bibnamefont{Frano}},
  \bibinfo{author}{\bibfnamefont{M.~M.} \bibnamefont{Yee}},
  \bibinfo{author}{\bibfnamefont{Y.}~\bibnamefont{Yoshida}},
  \bibinfo{author}{\bibfnamefont{H.}~\bibnamefont{Eisaki}},
  \bibinfo{author}{\bibfnamefont{E.}~\bibnamefont{Schierle}},
  \bibinfo{author}{\bibfnamefont{E.}~\bibnamefont{Weschke}},
  \bibinfo{author}{\bibfnamefont{R.}~\bibnamefont{Sutarto}},
  \bibinfo{author}{\bibfnamefont{F.}~\bibnamefont{He}},
  \bibinfo{author}{\bibfnamefont{A.}~\bibnamefont{Soumyanarayanan}},
  \bibnamefont{et~al.}, \bibinfo{journal}{Science}
  \textbf{\bibinfo{volume}{343}}, \bibinfo{pages}{390} (\bibinfo{year}{2014}),
  \urlprefix\url{https://doi.org/10.1126/science.1242996}.

\bibitem[{\citenamefont{Janod and \emph{et al.}}(2015)}]{Janod2015}
\bibinfo{author}{\bibfnamefont{E.}~\bibnamefont{Janod}} \bibnamefont{and}
  \bibinfo{author}{\bibnamefont{\emph{et al.}}}, \bibinfo{journal}{Adv. Funct.
  Mater.} \textbf{\bibinfo{volume}{25}}, \bibinfo{pages}{6287}
  (\bibinfo{year}{2015}).

\bibitem[{\citenamefont{Calandra}(2018)}]{PhysRevLett.121.026401}
\bibinfo{author}{\bibfnamefont{M.}~\bibnamefont{Calandra}},
  \bibinfo{journal}{Phys. Rev. Lett.} \textbf{\bibinfo{volume}{121}},
  \bibinfo{pages}{026401} (\bibinfo{year}{2018}),
  \urlprefix\url{https://link.aps.org/doi/10.1103/PhysRevLett.121.026401}.

\bibitem[{\citenamefont{Tresca and Calandra}(2019)}]{Tresca2019.NbS2}
\bibinfo{author}{\bibfnamefont{C.}~\bibnamefont{Tresca}} \bibnamefont{and}
  \bibinfo{author}{\bibfnamefont{M.}~\bibnamefont{Calandra}},
  \bibinfo{journal}{2D Materials} \textbf{\bibinfo{volume}{6}},
  \bibinfo{pages}{035041} (\bibinfo{year}{2019}),
  \urlprefix\url{https://doi.org/10.1088/2053-1583/ab23c0}.

\bibitem[{\citenamefont{Hubbard}(1963)}]{Hubbard1963}
\bibinfo{author}{\bibfnamefont{J.}~\bibnamefont{Hubbard}},
  \bibinfo{journal}{Proc. R. Soc. London} \textbf{\bibinfo{volume}{Ser. A276}},
  \bibinfo{pages}{238} (\bibinfo{year}{1963}).

\bibitem[{\citenamefont{Balents and \emph{et al.}}(2010)}]{Balents2010}
\bibinfo{author}{\bibfnamefont{L.}~\bibnamefont{Balents}} \bibnamefont{and}
  \bibinfo{author}{\bibnamefont{\emph{et al.}}}, \bibinfo{journal}{Nature}
  \textbf{\bibinfo{volume}{464}}, \bibinfo{pages}{199} (\bibinfo{year}{2010}).

\bibitem[{\citenamefont{Santoro et~al.}(1999)\citenamefont{Santoro, Scandolo,
  and Tosatti}}]{Santoro}
\bibinfo{author}{\bibfnamefont{G.}~\bibnamefont{Santoro}},
  \bibinfo{author}{\bibfnamefont{S.}~\bibnamefont{Scandolo}}, \bibnamefont{and}
  \bibinfo{author}{\bibfnamefont{E.}~\bibnamefont{Tosatti}},
  \bibinfo{journal}{Phys. Rev. B} \textbf{\bibinfo{volume}{59}},
  \bibinfo{pages}{1891} (\bibinfo{year}{1999}).

\bibitem[{\citenamefont{Profeta and Tosatti}(2007)}]{PhysRevLett.98.086401}
\bibinfo{author}{\bibfnamefont{G.}~\bibnamefont{Profeta}} \bibnamefont{and}
  \bibinfo{author}{\bibfnamefont{E.}~\bibnamefont{Tosatti}},
  \bibinfo{journal}{Phys. Rev. Lett.} \textbf{\bibinfo{volume}{98}},
  \bibinfo{pages}{086401} (\bibinfo{year}{2007}),
  \urlprefix\url{http://link.aps.org/doi/10.1103/PhysRevLett.98.086401}.

\bibitem[{\citenamefont{Hansmann
  et~al.}(2013{\natexlab{a}})\citenamefont{Hansmann, Ayral, Vaugier, Werner,
  and Biermann}}]{Hansmann2013a}
\bibinfo{author}{\bibfnamefont{P.}~\bibnamefont{Hansmann}},
  \bibinfo{author}{\bibfnamefont{T.}~\bibnamefont{Ayral}},
  \bibinfo{author}{\bibfnamefont{L.}~\bibnamefont{Vaugier}},
  \bibinfo{author}{\bibfnamefont{P.}~\bibnamefont{Werner}}, \bibnamefont{and}
  \bibinfo{author}{\bibfnamefont{S.}~\bibnamefont{Biermann}},
  \bibinfo{journal}{J.Phys.: Cond. Matter} \textbf{\bibinfo{volume}{25}},
  \bibinfo{pages}{094005} (\bibinfo{year}{2013}{\natexlab{a}}).

\bibitem[{\citenamefont{Tresca and \emph{et al.}}(2018)}]{Tresca2018}
\bibinfo{author}{\bibfnamefont{C.}~\bibnamefont{Tresca}} \bibnamefont{and}
  \bibinfo{author}{\bibnamefont{\emph{et al.}}}, \bibinfo{journal}{Phys. Rev.
  Lett.} \textbf{\bibinfo{volume}{120}}, \bibinfo{pages}{196402}
  (\bibinfo{year}{2018}).

\bibitem[{\citenamefont{Tresca and Calandra}(2021)}]{Tresca2019}
\bibinfo{author}{\bibfnamefont{C.}~\bibnamefont{Tresca}} \bibnamefont{and}
  \bibinfo{author}{\bibfnamefont{M.}~\bibnamefont{Calandra}},
  \bibinfo{journal}{Phys. Rev. B} \textbf{\bibinfo{volume}{104}},
  \bibinfo{pages}{045126} (\bibinfo{year}{2021}).

\bibitem[{\citenamefont{Brun et~al.}(2017)\citenamefont{Brun, Cren, and
  Roditchev}}]{Brun2017}
\bibinfo{author}{\bibfnamefont{C.}~\bibnamefont{Brun}},
  \bibinfo{author}{\bibfnamefont{T.}~\bibnamefont{Cren}}, \bibnamefont{and}
  \bibinfo{author}{\bibfnamefont{D.}~\bibnamefont{Roditchev}},
  \bibinfo{journal}{Supercond. Sci. Technol.} \textbf{\bibinfo{volume}{30}},
  \bibinfo{pages}{013003} (\bibinfo{year}{2017}).

\bibitem[{\citenamefont{Profeta et~al.}(2000)\citenamefont{Profeta, Continenza,
  Ottaviano, Mannstadt, and Freeman}}]{PhysRevB.62.1556}
\bibinfo{author}{\bibfnamefont{G.}~\bibnamefont{Profeta}},
  \bibinfo{author}{\bibfnamefont{A.}~\bibnamefont{Continenza}},
  \bibinfo{author}{\bibfnamefont{L.}~\bibnamefont{Ottaviano}},
  \bibinfo{author}{\bibfnamefont{W.}~\bibnamefont{Mannstadt}},
  \bibnamefont{and} \bibinfo{author}{\bibfnamefont{A.~J.}
  \bibnamefont{Freeman}}, \bibinfo{journal}{Phys. Rev. B}
  \textbf{\bibinfo{volume}{62}}, \bibinfo{pages}{1556} (\bibinfo{year}{2000}),
  \urlprefix\url{https://link.aps.org/doi/10.1103/PhysRevB.62.1556}.

\bibitem[{\citenamefont{Li et~al.}(2011)\citenamefont{Li, Laubach, Fleszar, and
  Hanke}}]{Li2011}
\bibinfo{author}{\bibfnamefont{G.}~\bibnamefont{Li}},
  \bibinfo{author}{\bibfnamefont{M.}~\bibnamefont{Laubach}},
  \bibinfo{author}{\bibfnamefont{A.}~\bibnamefont{Fleszar}}, \bibnamefont{and}
  \bibinfo{author}{\bibfnamefont{W.}~\bibnamefont{Hanke}},
  \bibinfo{journal}{Phys. Rev. B} \textbf{\bibinfo{volume}{83}},
  \bibinfo{pages}{041104} (\bibinfo{year}{2011}).

\bibitem[{\citenamefont{Hansmann
  et~al.}(2013{\natexlab{b}})\citenamefont{Hansmann, Ayral, Vaugier, Werner,
  and Biermann}}]{Hansmann2013b}
\bibinfo{author}{\bibfnamefont{P.}~\bibnamefont{Hansmann}},
  \bibinfo{author}{\bibfnamefont{T.}~\bibnamefont{Ayral}},
  \bibinfo{author}{\bibfnamefont{L.}~\bibnamefont{Vaugier}},
  \bibinfo{author}{\bibfnamefont{P.}~\bibnamefont{Werner}}, \bibnamefont{and}
  \bibinfo{author}{\bibfnamefont{S.}~\bibnamefont{Biermann}},
  \bibinfo{journal}{Phys. Rev. Lett.} \textbf{\bibinfo{volume}{110}},
  \bibinfo{pages}{166401} (\bibinfo{year}{2013}{\natexlab{b}}),
  \urlprefix\url{http://link.aps.org/doi/10.1103/PhysRevLett.110.166401}.

\bibitem[{\citenamefont{Cococcioni and
  de~Gironcoli}(2005)}]{PhysRevB.71.035105}
\bibinfo{author}{\bibfnamefont{M.}~\bibnamefont{Cococcioni}} \bibnamefont{and}
  \bibinfo{author}{\bibfnamefont{S.}~\bibnamefont{de~Gironcoli}},
  \bibinfo{journal}{Phys. Rev. B} \textbf{\bibinfo{volume}{71}},
  \bibinfo{pages}{035105} (\bibinfo{year}{2005}),
  \urlprefix\url{http://link.aps.org/doi/10.1103/PhysRevB.71.035105}.

\bibitem[{\citenamefont{Heyd et~al.}(2003)\citenamefont{Heyd, Scuseria, and
  Ernzerhof}}]{doi:10.1063/1.1564060}
\bibinfo{author}{\bibfnamefont{J.}~\bibnamefont{Heyd}},
  \bibinfo{author}{\bibfnamefont{G.~E.} \bibnamefont{Scuseria}},
  \bibnamefont{and}
  \bibinfo{author}{\bibfnamefont{M.}~\bibnamefont{Ernzerhof}},
  \bibinfo{journal}{The Journal of Chemical Physics}
  \textbf{\bibinfo{volume}{118}}, \bibinfo{pages}{8207} (\bibinfo{year}{2003}),
  \urlprefix\url{https://doi.org/10.1063/1.1564060}.

\bibitem[{\citenamefont{Heyd et~al.}(2006)\citenamefont{Heyd, Scuseria, and
  Ernzerhof}}]{doi:10.1063/1.2204597}
\bibinfo{author}{\bibfnamefont{J.}~\bibnamefont{Heyd}},
  \bibinfo{author}{\bibfnamefont{G.~E.} \bibnamefont{Scuseria}},
  \bibnamefont{and}
  \bibinfo{author}{\bibfnamefont{M.}~\bibnamefont{Ernzerhof}},
  \bibinfo{journal}{The Journal of Chemical Physics}
  \textbf{\bibinfo{volume}{124}}, \bibinfo{pages}{219906}
  (\bibinfo{year}{2006}), \urlprefix\url{https://doi.org/10.1063/1.2204597}.

\bibitem[{\citenamefont{Mascaraque and \emph{et al.}}(1999)}]{Mascaraque1999}
\bibinfo{author}{\bibfnamefont{A.}~\bibnamefont{Mascaraque}} \bibnamefont{and}
  \bibinfo{author}{\bibnamefont{\emph{et al.}}}, \bibinfo{journal}{Phys. Rev.
  Lett.} \textbf{\bibinfo{volume}{82}}, \bibinfo{pages}{2524}
  (\bibinfo{year}{1999}).

\bibitem[{\citenamefont{Zhang and \emph{et al.}}(1999)}]{Zhang1999}
\bibinfo{author}{\bibfnamefont{J.}~\bibnamefont{Zhang}} \bibnamefont{and}
  \bibinfo{author}{\bibnamefont{\emph{et al.}}}, \bibinfo{journal}{Phys. Rev.
  B} \textbf{\bibinfo{volume}{60}}, \bibinfo{pages}{2860}
  (\bibinfo{year}{1999}).

\bibitem[{\citenamefont{Bunk and \emph{et al.}}(1999)}]{Bunk1999}
\bibinfo{author}{\bibfnamefont{O.}~\bibnamefont{Bunk}} \bibnamefont{and}
  \bibinfo{author}{\bibnamefont{\emph{et al.}}}, \bibinfo{journal}{Phys. Rev.
  Lett.} \textbf{\bibinfo{volume}{83}}, \bibinfo{pages}{2226}
  (\bibinfo{year}{1999}).

\bibitem[{\citenamefont{Custance and \emph{et al.}}(2001)}]{Custance2001b}
\bibinfo{author}{\bibfnamefont{O.}~\bibnamefont{Custance}} \bibnamefont{and}
  \bibinfo{author}{\bibnamefont{\emph{et al.}}}, \bibinfo{journal}{Surf. Sci.}
  \textbf{\bibinfo{volume}{482}}, \bibinfo{pages}{1399} (\bibinfo{year}{2001}).

\bibitem[{\citenamefont{Brihuega
  et~al.}(2005{\natexlab{b}})\citenamefont{Brihuega, Custance, P\'erez, and
  G\'omez-Rodr\'igues}}]{Brihuega2005}
\bibinfo{author}{\bibfnamefont{I.}~\bibnamefont{Brihuega}},
  \bibinfo{author}{\bibfnamefont{O.}~\bibnamefont{Custance}},
  \bibinfo{author}{\bibfnamefont{R.}~\bibnamefont{P\'erez}}, \bibnamefont{and}
  \bibinfo{author}{\bibfnamefont{J.}~\bibnamefont{G\'omez-Rodr\'igues}},
  \bibinfo{journal}{Phys.Rev.Lett.} \textbf{\bibinfo{volume}{94}},
  \bibinfo{pages}{046101} (\bibinfo{year}{2005}{\natexlab{b}}).

\bibitem[{\citenamefont{Adler and \emph{et al.}}(2019)}]{Adler2019}
\bibinfo{author}{\bibfnamefont{F.}~\bibnamefont{Adler}} \bibnamefont{and}
  \bibinfo{author}{\bibnamefont{\emph{et al.}}}, \bibinfo{journal}{Phys. Rev.
  Lett.} \textbf{\bibinfo{volume}{123}}, \bibinfo{pages}{086401}
  (\bibinfo{year}{2019}).

\bibitem[{\citenamefont{Brihuega
  et~al.}(2007{\natexlab{a}})\citenamefont{Brihuega, Custance, Ugeda, and
  G\'omez-Rodr\'igues}}]{Brihuega2007}
\bibinfo{author}{\bibfnamefont{I.}~\bibnamefont{Brihuega}},
  \bibinfo{author}{\bibfnamefont{O.}~\bibnamefont{Custance}},
  \bibinfo{author}{\bibfnamefont{M.}~\bibnamefont{Ugeda}}, \bibnamefont{and}
  \bibinfo{author}{\bibfnamefont{J.}~\bibnamefont{G\'omez-Rodr\'igues}},
  \bibinfo{journal}{Phys.Rev.Lett.} \textbf{\bibinfo{volume}{75}},
  \bibinfo{pages}{155411} (\bibinfo{year}{2007}{\natexlab{a}}).

\bibitem[{\citenamefont{Salamon and Jaime}(2001)}]{Salamon2001}
\bibinfo{author}{\bibfnamefont{M.~B.} \bibnamefont{Salamon}} \bibnamefont{and}
  \bibinfo{author}{\bibfnamefont{M.}~\bibnamefont{Jaime}},
  \bibinfo{journal}{Rev. Mod. Phys.} \textbf{\bibinfo{volume}{73}},
  \bibinfo{pages}{583} (\bibinfo{year}{2001}),
  \urlprefix\url{https://link.aps.org/doi/10.1103/RevModPhys.73.583}.

\bibitem[{\citenamefont{Cao and Schlottmann}(2018)}]{Cao2018}
\bibinfo{author}{\bibfnamefont{G.}~\bibnamefont{Cao}} \bibnamefont{and}
  \bibinfo{author}{\bibfnamefont{P.}~\bibnamefont{Schlottmann}},
  \bibinfo{journal}{Reports on Progress in Physics}
  \textbf{\bibinfo{volume}{81}}, \bibinfo{pages}{042502}
  (\bibinfo{year}{2018}),
  \urlprefix\url{https://doi.org/10.1088/1361-6633/aaa979}.

\bibitem[{\citenamefont{Wosnitza}(2007)}]{Wosnitza2007}
\bibinfo{author}{\bibfnamefont{J.}~\bibnamefont{Wosnitza}},
  \bibinfo{journal}{J. Low Temp. Phys.} \textbf{\bibinfo{volume}{146}},
  \bibinfo{pages}{641} (\bibinfo{year}{2007}),
  \urlprefix\url{https://doi.org/10.1007/s10909-006-9282-9}.

\bibitem[{\citenamefont{Lebed}(2008)}]{bookorganics}
\bibinfo{author}{\bibfnamefont{A.}~\bibnamefont{Lebed}},
  \emph{\bibinfo{title}{The Physics of Organic Superconductors and Conductors}}
  (\bibinfo{publisher}{Springer Berlin, Heidelberg}, \bibinfo{year}{2008}),
  ISBN \bibinfo{isbn}{978-3-540-76672-8},
  \urlprefix\url{https://doi.org/10.1007/978-3-540-76672-8}.

\bibitem[{\citenamefont{Andrei and MacDonald}(2020)}]{bilayergraphene}
\bibinfo{author}{\bibfnamefont{E.}~\bibnamefont{Andrei}} \bibnamefont{and}
  \bibinfo{author}{\bibfnamefont{A.}~\bibnamefont{MacDonald}},
  \bibinfo{journal}{Nat. Mater.} \textbf{\bibinfo{volume}{19}},
  \bibinfo{pages}{1265} (\bibinfo{year}{2020}),
  \urlprefix\url{https://doi.org/10.1038/s41563-020-00840-0}.

\bibitem[{\citenamefont{Tersoff and Hamann}(1985)}]{TersoffHamann}
\bibinfo{author}{\bibfnamefont{J.}~\bibnamefont{Tersoff}} \bibnamefont{and}
  \bibinfo{author}{\bibfnamefont{D.~R.} \bibnamefont{Hamann}},
  \bibinfo{journal}{Phys. Rev. B} \textbf{\bibinfo{volume}{31}},
  \bibinfo{pages}{805} (\bibinfo{year}{1985}),
  \urlprefix\url{https://link.aps.org/doi/10.1103/PhysRevB.31.805}.

\bibitem[{\citenamefont{Li et~al.}(1997)\citenamefont{Li, Schneider, and
  Berndt}}]{Li1997}
\bibinfo{author}{\bibfnamefont{J.}~\bibnamefont{Li}},
  \bibinfo{author}{\bibfnamefont{W.-D.} \bibnamefont{Schneider}},
  \bibnamefont{and} \bibinfo{author}{\bibfnamefont{R.}~\bibnamefont{Berndt}},
  \bibinfo{journal}{Phys. Rev. B} \textbf{\bibinfo{volume}{56}},
  \bibinfo{pages}{7656} (\bibinfo{year}{1997}),
  \urlprefix\url{https://link.aps.org/doi/10.1103/PhysRevB.56.7656}.

\bibitem[{\citenamefont{Ziegler and \emph{et al.}}(2009)}]{Ziegler2009}
\bibinfo{author}{\bibfnamefont{M.}~\bibnamefont{Ziegler}} \bibnamefont{and}
  \bibinfo{author}{\bibnamefont{\emph{et al.}}}, \bibinfo{journal}{Phys. Rev.
  B} \textbf{\bibinfo{volume}{80}}, \bibinfo{pages}{125402}
  (\bibinfo{year}{2009}).

\bibitem[{\citenamefont{Lu and \emph{et al.}}(2003)}]{Lu2003}
\bibinfo{author}{\bibfnamefont{X.}~\bibnamefont{Lu}} \bibnamefont{and}
  \bibinfo{author}{\bibnamefont{\emph{et al.}}}, \bibinfo{journal}{Phys. Rev.
  Lett.} \textbf{\bibinfo{volume}{90}}, \bibinfo{pages}{096802}
  (\bibinfo{year}{2003}).

\bibitem[{\citenamefont{Reecht and \emph{et al.}}(2017)}]{Reecht2017}
\bibinfo{author}{\bibfnamefont{G.}~\bibnamefont{Reecht}} \bibnamefont{and}
  \bibinfo{author}{\bibnamefont{\emph{et al.}}}, \bibinfo{journal}{New J.
  Phys.} \textbf{\bibinfo{volume}{19}}, \bibinfo{pages}{113033}
  (\bibinfo{year}{2017}).

\bibitem[{\citenamefont{Fischer et~al.}(2007)\citenamefont{Fischer, Kugler,
  Maggio-Aprile, Berthod, and Renner}}]{Fischer}
\bibinfo{author}{\bibfnamefont{O.}~\bibnamefont{Fischer}},
  \bibinfo{author}{\bibfnamefont{M.}~\bibnamefont{Kugler}},
  \bibinfo{author}{\bibfnamefont{I.}~\bibnamefont{Maggio-Aprile}},
  \bibinfo{author}{\bibfnamefont{C.}~\bibnamefont{Berthod}}, \bibnamefont{and}
  \bibinfo{author}{\bibfnamefont{C.}~\bibnamefont{Renner}},
  \bibinfo{journal}{Rev. Mod. Phys.} \textbf{\bibinfo{volume}{79}},
  \bibinfo{pages}{353} (\bibinfo{year}{2007}),
  \urlprefix\url{https://link.aps.org/doi/10.1103/RevModPhys.79.353}.

\bibitem[{\citenamefont{Simon et~al.}(2011)\citenamefont{Simon, Bena, Vonau,
  Cranney, and Aubel}}]{SimonFTSTS}
\bibinfo{author}{\bibfnamefont{L.}~\bibnamefont{Simon}},
  \bibinfo{author}{\bibfnamefont{C.}~\bibnamefont{Bena}},
  \bibinfo{author}{\bibfnamefont{F.}~\bibnamefont{Vonau}},
  \bibinfo{author}{\bibfnamefont{M.}~\bibnamefont{Cranney}}, \bibnamefont{and}
  \bibinfo{author}{\bibfnamefont{D.}~\bibnamefont{Aubel}},
  \bibinfo{journal}{Journal of Physics D: Applied Physics}
  \textbf{\bibinfo{volume}{44}}, \bibinfo{pages}{464010}
  (\bibinfo{year}{2011}),
  \urlprefix\url{https://doi.org/10.1088/0022-3727/44/46/464010}.

\bibitem[{\citenamefont{Chen et~al.}(2017)\citenamefont{Chen, Cheng, and
  Wu}}]{Chen2017}
\bibinfo{author}{\bibfnamefont{L.}~\bibnamefont{Chen}},
  \bibinfo{author}{\bibfnamefont{P.}~\bibnamefont{Cheng}}, \bibnamefont{and}
  \bibinfo{author}{\bibfnamefont{K.}~\bibnamefont{Wu}},
  \bibinfo{journal}{Journal of Physics: Condensed Matter}
  \textbf{\bibinfo{volume}{29}}, \bibinfo{pages}{103001}
  (\bibinfo{year}{2017}),
  \urlprefix\url{https://doi.org/10.1088/1361-648x/aa54da}.

\bibitem[{\citenamefont{Lin et~al.}(2020)\citenamefont{Lin, Kawakami, Arafune,
  Minamitani, and Takagi}}]{Lin2020}
\bibinfo{author}{\bibfnamefont{C.-L.} \bibnamefont{Lin}},
  \bibinfo{author}{\bibfnamefont{N.}~\bibnamefont{Kawakami}},
  \bibinfo{author}{\bibfnamefont{R.}~\bibnamefont{Arafune}},
  \bibinfo{author}{\bibfnamefont{E.}~\bibnamefont{Minamitani}},
  \bibnamefont{and} \bibinfo{author}{\bibfnamefont{N.}~\bibnamefont{Takagi}},
  \bibinfo{journal}{J. Phys.: Condens. Matter} \textbf{\bibinfo{volume}{32}},
  \bibinfo{pages}{243001} (\bibinfo{year}{2020}).

\bibitem[{\citenamefont{Ming and \emph{et al.}}(2017)}]{Ming}
\bibinfo{author}{\bibfnamefont{F.}~\bibnamefont{Ming}} \bibnamefont{and}
  \bibinfo{author}{\bibnamefont{\emph{et al.}}}, \bibinfo{journal}{Phys. Rev.
  Lett.} \textbf{\bibinfo{volume}{119}}, \bibinfo{pages}{266802}
  (\bibinfo{year}{2017}).

\bibitem[{\citenamefont{Giannozzi and {\it et al.}}(2009)}]{QEcode}
\bibinfo{author}{\bibfnamefont{P.}~\bibnamefont{Giannozzi}} \bibnamefont{and}
  \bibinfo{author}{\bibnamefont{{\it et al.}}}, \bibinfo{journal}{Journal of
  Physics: Condensed Matter} \textbf{\bibinfo{volume}{21}},
  \bibinfo{pages}{395502} (\bibinfo{year}{2009}),
  \urlprefix\url{http://stacks.iop.org/0953-8984/21/i=39/a=395502}.

\bibitem[{\citenamefont{Giannozzi et~al.}(2017)\citenamefont{Giannozzi,
  Andreussi, Brumme, Bunau, Nardelli, Calandra, Car, Cavazzoni, Ceresoli,
  Cococcioni et~al.}}]{QE-2017}
\bibinfo{author}{\bibfnamefont{P.}~\bibnamefont{Giannozzi}},
  \bibinfo{author}{\bibfnamefont{O.}~\bibnamefont{Andreussi}},
  \bibinfo{author}{\bibfnamefont{T.}~\bibnamefont{Brumme}},
  \bibinfo{author}{\bibfnamefont{O.}~\bibnamefont{Bunau}},
  \bibinfo{author}{\bibfnamefont{M.~B.} \bibnamefont{Nardelli}},
  \bibinfo{author}{\bibfnamefont{M.}~\bibnamefont{Calandra}},
  \bibinfo{author}{\bibfnamefont{R.}~\bibnamefont{Car}},
  \bibinfo{author}{\bibfnamefont{C.}~\bibnamefont{Cavazzoni}},
  \bibinfo{author}{\bibfnamefont{D.}~\bibnamefont{Ceresoli}},
  \bibinfo{author}{\bibfnamefont{M.}~\bibnamefont{Cococcioni}},
  \bibnamefont{et~al.}, \bibinfo{journal}{Journal of Physics: Condensed Matter}
  \textbf{\bibinfo{volume}{29}}, \bibinfo{pages}{465901}
  (\bibinfo{year}{2017}),
  \urlprefix\url{http://stacks.iop.org/0953-8984/29/i=46/a=465901}.

\bibitem[{\citenamefont{Dovesi et~al.}(2018{\natexlab{a}})\citenamefont{Dovesi,
  Erba, Orlando, Zicovich-Wilson, Civalleri, Maschio, Rérat, Casassa, Baima,
  Salustro et~al.}}]{doi:10.1002/wcms.1360}
\bibinfo{author}{\bibfnamefont{R.}~\bibnamefont{Dovesi}},
  \bibinfo{author}{\bibfnamefont{A.}~\bibnamefont{Erba}},
  \bibinfo{author}{\bibfnamefont{R.}~\bibnamefont{Orlando}},
  \bibinfo{author}{\bibfnamefont{C.~M.} \bibnamefont{Zicovich-Wilson}},
  \bibinfo{author}{\bibfnamefont{B.}~\bibnamefont{Civalleri}},
  \bibinfo{author}{\bibfnamefont{L.}~\bibnamefont{Maschio}},
  \bibinfo{author}{\bibfnamefont{M.}~\bibnamefont{Rérat}},
  \bibinfo{author}{\bibfnamefont{S.}~\bibnamefont{Casassa}},
  \bibinfo{author}{\bibfnamefont{J.}~\bibnamefont{Baima}},
  \bibinfo{author}{\bibfnamefont{S.}~\bibnamefont{Salustro}},
  \bibnamefont{et~al.}, \bibinfo{journal}{Wiley Interdisciplinary Reviews:
  Computational Molecular Science} \textbf{\bibinfo{volume}{8}},
  \bibinfo{pages}{e1360} (\bibinfo{year}{2018}{\natexlab{a}}),
  \urlprefix\url{https://onlinelibrary.wiley.com/doi/abs/10.1002/wcms.1360}.

\bibitem[{\citenamefont{Dovesi et~al.}(2018{\natexlab{b}})\citenamefont{Dovesi,
  Saunders, Roetti, Orlando, Zicovich-Wilson, Pascale, Civalleri, Doll,
  Harrison, Bush et~al.}}]{cryman}
\bibinfo{author}{\bibfnamefont{R.}~\bibnamefont{Dovesi}},
  \bibinfo{author}{\bibfnamefont{V.~R.} \bibnamefont{Saunders}},
  \bibinfo{author}{\bibfnamefont{C.}~\bibnamefont{Roetti}},
  \bibinfo{author}{\bibfnamefont{R.}~\bibnamefont{Orlando}},
  \bibinfo{author}{\bibfnamefont{C.~M.} \bibnamefont{Zicovich-Wilson}},
  \bibinfo{author}{\bibfnamefont{F.}~\bibnamefont{Pascale}},
  \bibinfo{author}{\bibfnamefont{B.}~\bibnamefont{Civalleri}},
  \bibinfo{author}{\bibfnamefont{K.}~\bibnamefont{Doll}},
  \bibinfo{author}{\bibfnamefont{N.~M.} \bibnamefont{Harrison}},
  \bibinfo{author}{\bibfnamefont{I.~J.} \bibnamefont{Bush}},
  \bibnamefont{et~al.}, \bibinfo{journal}{CRYSTAL17 User's Manual}
  (\bibinfo{year}{2018}{\natexlab{b}}),
  \urlprefix\url{https://www.crystal.unito.it/Manuals/crystal17.pdf}.

\bibitem[{\citenamefont{Monkhorst and Pack}(1976)}]{PhysRevB.13.5188}
\bibinfo{author}{\bibfnamefont{H.~J.} \bibnamefont{Monkhorst}}
  \bibnamefont{and} \bibinfo{author}{\bibfnamefont{J.~D.} \bibnamefont{Pack}},
  \bibinfo{journal}{Phys. Rev. B} \textbf{\bibinfo{volume}{13}},
  \bibinfo{pages}{5188} (\bibinfo{year}{1976}),
  \urlprefix\url{http://link.aps.org/doi/10.1103/PhysRevB.13.5188}.

\bibitem[{\citenamefont{Sophia et~al.}(2013)\citenamefont{Sophia, Baranek,
  Sarrazin, R{\'{e}}rat, and Dovesi}}]{Sophia2013}
\bibinfo{author}{\bibfnamefont{G.}~\bibnamefont{Sophia}},
  \bibinfo{author}{\bibfnamefont{P.}~\bibnamefont{Baranek}},
  \bibinfo{author}{\bibfnamefont{C.}~\bibnamefont{Sarrazin}},
  \bibinfo{author}{\bibfnamefont{M.}~\bibnamefont{R{\'{e}}rat}},
  \bibnamefont{and} \bibinfo{author}{\bibfnamefont{R.}~\bibnamefont{Dovesi}},
  \bibinfo{journal}{Phase Transitions} \textbf{\bibinfo{volume}{86}},
  \bibinfo{pages}{1069} (\bibinfo{year}{2013}),
  \urlprefix\url{https://doi.org/10.1080/01411594.2012.754442}.

\bibitem[{\citenamefont{Peralta et~al.}(2006)\citenamefont{Peralta, Heyd,
  Scuseria, and Martin}}]{PhysRevB.74.073101}
\bibinfo{author}{\bibfnamefont{J.~E.} \bibnamefont{Peralta}},
  \bibinfo{author}{\bibfnamefont{J.}~\bibnamefont{Heyd}},
  \bibinfo{author}{\bibfnamefont{G.~E.} \bibnamefont{Scuseria}},
  \bibnamefont{and} \bibinfo{author}{\bibfnamefont{R.~L.}
  \bibnamefont{Martin}}, \bibinfo{journal}{Phys. Rev. B}
  \textbf{\bibinfo{volume}{74}}, \bibinfo{pages}{073101}
  (\bibinfo{year}{2006}),
  \urlprefix\url{https://link.aps.org/doi/10.1103/PhysRevB.74.073101}.

\bibitem[{\citenamefont{Heyd et~al.}(2005)\citenamefont{Heyd, Peralta,
  Scuseria, and Martin}}]{doi:10.1063/1.2085170}
\bibinfo{author}{\bibfnamefont{J.}~\bibnamefont{Heyd}},
  \bibinfo{author}{\bibfnamefont{J.~E.} \bibnamefont{Peralta}},
  \bibinfo{author}{\bibfnamefont{G.~E.} \bibnamefont{Scuseria}},
  \bibnamefont{and} \bibinfo{author}{\bibfnamefont{R.~L.}
  \bibnamefont{Martin}}, \bibinfo{journal}{The Journal of Chemical Physics}
  \textbf{\bibinfo{volume}{123}}, \bibinfo{pages}{174101}
  (\bibinfo{year}{2005}), \urlprefix\url{https://doi.org/10.1063/1.2085170}.

\bibitem[{\citenamefont{Pernot et~al.}(2015)\citenamefont{Pernot, Civalleri,
  Presti, and Savin}}]{Pernot2015}
\bibinfo{author}{\bibfnamefont{P.}~\bibnamefont{Pernot}},
  \bibinfo{author}{\bibfnamefont{B.}~\bibnamefont{Civalleri}},
  \bibinfo{author}{\bibfnamefont{D.}~\bibnamefont{Presti}}, \bibnamefont{and}
  \bibinfo{author}{\bibfnamefont{A.}~\bibnamefont{Savin}},
  \bibinfo{journal}{The Journal of Physical Chemistry A}
  \textbf{\bibinfo{volume}{119}}, \bibinfo{pages}{5288} (\bibinfo{year}{2015}),
  \urlprefix\url{https://doi.org/10.1021/jp509980w}.

\bibitem[{\citenamefont{Peintinger et~al.}(2012)\citenamefont{Peintinger,
  Oliveira, and Bredow}}]{Peintinger2012}
\bibinfo{author}{\bibfnamefont{M.~F.} \bibnamefont{Peintinger}},
  \bibinfo{author}{\bibfnamefont{D.~V.} \bibnamefont{Oliveira}},
  \bibnamefont{and} \bibinfo{author}{\bibfnamefont{T.}~\bibnamefont{Bredow}},
  \bibinfo{journal}{Journal of Computational Chemistry}
  \textbf{\bibinfo{volume}{34}}, \bibinfo{pages}{451} (\bibinfo{year}{2012}),
  \urlprefix\url{https://doi.org/10.1002/jcc.23153}.

\bibitem[{\citenamefont{Mascaraque et~al.}(1998)\citenamefont{Mascaraque,
  Avila, Michel, and Asensio}}]{PhysRevB.57.14758}
\bibinfo{author}{\bibfnamefont{A.}~\bibnamefont{Mascaraque}},
  \bibinfo{author}{\bibfnamefont{J.}~\bibnamefont{Avila}},
  \bibinfo{author}{\bibfnamefont{E.~G.} \bibnamefont{Michel}},
  \bibnamefont{and} \bibinfo{author}{\bibfnamefont{M.~C.}
  \bibnamefont{Asensio}}, \bibinfo{journal}{Phys. Rev. B}
  \textbf{\bibinfo{volume}{57}}, \bibinfo{pages}{14758} (\bibinfo{year}{1998}),
  \urlprefix\url{https://link.aps.org/doi/10.1103/PhysRevB.57.14758}.

\bibitem[{\citenamefont{Mascaraque
  et~al.}(1999{\natexlab{a}})\citenamefont{Mascaraque, Avila, Asensio, and
  Michel}}]{MASCARAQUE1999337}
\bibinfo{author}{\bibfnamefont{A.}~\bibnamefont{Mascaraque}},
  \bibinfo{author}{\bibfnamefont{J.}~\bibnamefont{Avila}},
  \bibinfo{author}{\bibfnamefont{M.}~\bibnamefont{Asensio}}, \bibnamefont{and}
  \bibinfo{author}{\bibfnamefont{E.}~\bibnamefont{Michel}},
  \bibinfo{journal}{Surface Science} \textbf{\bibinfo{volume}{433-435}},
  \bibinfo{pages}{337 } (\bibinfo{year}{1999}{\natexlab{a}}), ISSN
  \bibinfo{issn}{0039-6028},
  \urlprefix\url{http://www.sciencedirect.com/science/article/pii/S0039602899001296}.

\bibitem[{\citenamefont{Brihuega
  et~al.}(2007{\natexlab{b}})\citenamefont{Brihuega, Custance, Ugeda, and
  G\'omez-Rodr\'{\i}guez}}]{PhysRevB.75.155411}
\bibinfo{author}{\bibfnamefont{I.}~\bibnamefont{Brihuega}},
  \bibinfo{author}{\bibfnamefont{O.}~\bibnamefont{Custance}},
  \bibinfo{author}{\bibfnamefont{M.~M.} \bibnamefont{Ugeda}}, \bibnamefont{and}
  \bibinfo{author}{\bibfnamefont{J.~M.} \bibnamefont{G\'omez-Rodr\'{\i}guez}},
  \bibinfo{journal}{Phys. Rev. B} \textbf{\bibinfo{volume}{75}},
  \bibinfo{pages}{155411} (\bibinfo{year}{2007}{\natexlab{b}}),
  \urlprefix\url{https://link.aps.org/doi/10.1103/PhysRevB.75.155411}.

\bibitem[{\citenamefont{Tejeda et~al.}(2007)\citenamefont{Tejeda, Cort{\'{e}}s,
  Lobo, Michel, and Mascaraque}}]{Tejeda_2007}
\bibinfo{author}{\bibfnamefont{A.}~\bibnamefont{Tejeda}},
  \bibinfo{author}{\bibfnamefont{R.}~\bibnamefont{Cort{\'{e}}s}},
  \bibinfo{author}{\bibfnamefont{J.}~\bibnamefont{Lobo}},
  \bibinfo{author}{\bibfnamefont{E.~G.} \bibnamefont{Michel}},
  \bibnamefont{and}
  \bibinfo{author}{\bibfnamefont{A.}~\bibnamefont{Mascaraque}},
  \bibinfo{journal}{Journal of Physics: Condensed Matter}
  \textbf{\bibinfo{volume}{19}}, \bibinfo{pages}{355008}
  (\bibinfo{year}{2007}),
  \urlprefix\url{https://doi.org/10.1088/0953-8984/19/35/355008}.

\bibitem[{\citenamefont{Cudazzo et~al.}(2008)\citenamefont{Cudazzo, Profeta,
  and Continenza}}]{Cudazzo2008747}
\bibinfo{author}{\bibfnamefont{P.}~\bibnamefont{Cudazzo}},
  \bibinfo{author}{\bibfnamefont{G.}~\bibnamefont{Profeta}}, \bibnamefont{and}
  \bibinfo{author}{\bibfnamefont{A.}~\bibnamefont{Continenza}},
  \bibinfo{journal}{Surface Science} \textbf{\bibinfo{volume}{602}},
  \bibinfo{pages}{747 } (\bibinfo{year}{2008}).

\bibitem[{\citenamefont{Mascaraque
  et~al.}(1999{\natexlab{b}})\citenamefont{Mascaraque, Avila, Alvarez, Asensio,
  Ferrer, and Michel}}]{PhysRevLett.82.2524}
\bibinfo{author}{\bibfnamefont{A.}~\bibnamefont{Mascaraque}},
  \bibinfo{author}{\bibfnamefont{J.}~\bibnamefont{Avila}},
  \bibinfo{author}{\bibfnamefont{J.}~\bibnamefont{Alvarez}},
  \bibinfo{author}{\bibfnamefont{M.~C.} \bibnamefont{Asensio}},
  \bibinfo{author}{\bibfnamefont{S.}~\bibnamefont{Ferrer}}, \bibnamefont{and}
  \bibinfo{author}{\bibfnamefont{E.~G.} \bibnamefont{Michel}},
  \bibinfo{journal}{Phys. Rev. Lett.} \textbf{\bibinfo{volume}{82}},
  \bibinfo{pages}{2524} (\bibinfo{year}{1999}{\natexlab{b}}),
  \urlprefix\url{https://link.aps.org/doi/10.1103/PhysRevLett.82.2524}.

\bibitem[{\citenamefont{Badrtdinov et~al.}(2016)\citenamefont{Badrtdinov,
  Nikolaev, Katsnelson, and Mazurenko}}]{PhysRevB.94.224418}
\bibinfo{author}{\bibfnamefont{D.~I.} \bibnamefont{Badrtdinov}},
  \bibinfo{author}{\bibfnamefont{S.~A.} \bibnamefont{Nikolaev}},
  \bibinfo{author}{\bibfnamefont{M.~I.} \bibnamefont{Katsnelson}},
  \bibnamefont{and} \bibinfo{author}{\bibfnamefont{V.~V.}
  \bibnamefont{Mazurenko}}, \bibinfo{journal}{Phys. Rev. B}
  \textbf{\bibinfo{volume}{94}}, \bibinfo{pages}{224418}
  (\bibinfo{year}{2016}),
  \urlprefix\url{https://link.aps.org/doi/10.1103/PhysRevB.94.224418}.

\bibitem[{\citenamefont{Lee et~al.}(2013)\citenamefont{Lee, Kim, and
  Cho}}]{Lee2013}
\bibinfo{author}{\bibfnamefont{J.-H.} \bibnamefont{Lee}},
  \bibinfo{author}{\bibfnamefont{H.-J.} \bibnamefont{Kim}}, \bibnamefont{and}
  \bibinfo{author}{\bibfnamefont{J.-H.} \bibnamefont{Cho}},
  \bibinfo{journal}{Phys. Rev. Lett.} \textbf{\bibinfo{volume}{111}},
  \bibinfo{pages}{106403} (\bibinfo{year}{2013}),
  \urlprefix\url{https://link.aps.org/doi/10.1103/PhysRevLett.111.106403}.

\bibitem[{\citenamefont{Lee et~al.}(2014)\citenamefont{Lee, Ren, Jia, and
  Cho}}]{Lee2014}
\bibinfo{author}{\bibfnamefont{J.-H.} \bibnamefont{Lee}},
  \bibinfo{author}{\bibfnamefont{X.-Y.} \bibnamefont{Ren}},
  \bibinfo{author}{\bibfnamefont{Y.}~\bibnamefont{Jia}}, \bibnamefont{and}
  \bibinfo{author}{\bibfnamefont{J.-H.} \bibnamefont{Cho}},
  \bibinfo{journal}{Phys. Rev. B} \textbf{\bibinfo{volume}{90}},
  \bibinfo{pages}{125439} (\bibinfo{year}{2014}),
  \urlprefix\url{https://link.aps.org/doi/10.1103/PhysRevB.90.125439}.

\end{thebibliography}

\end{document}